\documentclass[iop,letterpaper]{emulateapj}

\usepackage{graphicx, subfigure, float, amsmath, 
            natbib, epstopdf, grffile}

\shorttitle{Stellar Activity on GJ 1245 A and B}
\shortauthors{Lurie et al.}

\begin{document}


\title{Kepler Flares III: Stellar Activity on GJ 1245 A and B}

\author{John C. Lurie\altaffilmark{1}, James
  R. A. Davenport\altaffilmark{1}, Suzanne L. Hawley\altaffilmark{1},
  Tessa D. Wilkinson\altaffilmark{1}, John P. Wisniewski\altaffilmark{2},\\
  Adam F. Kowalski\altaffilmark{3}, and Leslie Hebb\altaffilmark{4}}

\altaffiltext{1}{Department of Astronomy, University of Washington, Seattle, WA 98195, USA; \textbf{lurie@uw.edu}}
\altaffiltext{2}{HL Dodge Department of Physics \& Astronomy, University of Oklahoma, Norman, OK 73019, USA}
\altaffiltext{3}{NASA Goddard Space Flight Center, Greenbelt, MD 20771, USA}
\altaffiltext{4}{Department of Physics, Hobart and William Smith Colleges, Geneva, NY 14456, USA}

\begin{abstract}

We present the flare occurrence rates and starspot evolution for GJ
1245 A and B, two active M5 stars, based on nine months of Kepler
short cadence observations, and four years of nearly continuous long
cadence observations. The A component is separated from the B
component by 7\arcsec, and the stars are not resolved in the Kepler
pipeline processing due to Kepler's large plate scale of
4\arcsec/pixel. Analyzing the target pixel data, we have generated
separate light curves for components A and B using the PyKE pixel
response function modeling procedures, and note the effects of CCD
saturation and non-linear response to high energy flares. In our
sample, GJ 1245A and B exhibit an average of 3.0 and 2.6 flares per
day, respectively. We introduce a new metric,
$L_{fl}/L_{\mathrm{Kp}}$, to compare the flare rates between stars,
and discuss this in the context of GJ 1245 A and B. Both stars exhibit
starspot features that evolve on long time scales, with the slower
rotating B component showing evidence of differential
rotation. Intriguingly, the angular separation between the A and B
component photocenters decreases during the four years of observations
in a manner consistent with a shift in the position of the A
photocenter due to the orbit of its unseen M8 companion (GJ 1245C),
which is $\sim$94\% less bright. Among the most detailed photometric
studies of fully convective M dwarfs in a multiple system, these
results provide an important constraint on stellar
age-rotation-activity models.

\end{abstract}


\section{Introduction}
\label{sec:intro}

The magnetic fields of M dwarfs manifest themselves in several
observable ways. These include flares (e.g, \citealt{Lacy1976}),
starspots (e.g., \citealt{Contadakis1995}; \citealt{McQuillan2014}),
chromospheric H$\alpha$ emission (e.g., \citealt{Hawley1996}), and
X-ray emission (e.g., \citealt{Gudel2004}). There has been a
longstanding effort to tie such observables to the internal magnetic
dynamo, and to disentangle the interdependent effects of stellar mass,
age, and rotation rate. In the age-rotation-activity paradigm (e.g.,
\citealt{Skumanich1972}), activity depends on rotation rate, which in
turn depends on age. Complicating matters, M dwarfs become fully
convective at approximately type M4 \citep{Chabrier1997}, and thus do
not have a Solar-like dynamo. While models indicate that activity in
fully convective stars depends on rotation rate
\citep{Dobler2006,Browning2008}, there is observational evidence for a
rotation threshold \citep{Delfosse1998,Mohanty2003,Browning2010},
above which activity no longer correlates with rotation rate.

Because they are coeval, stars in multiple systems provide a control
for age, and are test cases for the age-rotation-activity
paradigm. Among the nearest (4.5 pc; \citealt{vanAltena1995}) and
brightest M dwarfs in the Kepler dataset, the GJ 1245 system is
comprised of two active M5 components (A and B), and an M8 companion
(C) to A. The spectral types reported here are those in
\citet{Hawley2014}, hereafter referred to as Paper 1.  At $\sim$3
magnitudes fainter ($94\%$ less bright) than component A, component C
does not contribute significantly to the total quiescent flux in the
Kepler bandpass.  Kepler has observed flares on stars as late as L1
\citep{Gizis2013}, and it is possible that a flare on component C
could be detected and mistakenly assigned to component A. We discuss
the contribution of the C component to our flare sample uncertainties
in $\S$\ref{sec:flares}. For simplicity, we refer to the properties of
the A component individually unless otherwise noted, but the Kepler
observations presented here are of the combined flux from the A and C
components.

The A and C components are separated by $\sim$0\farcs6 (2.7 AU)
\citep{Dieterich2012} with an orbital period of $\sim$15 years
\citep{Harrington1990}, while the AC and B components are separated by
$\sim$7{\arcsec} (32 AU), with an estimated orbital period of 330
years assuming a circular orbit and a total system mass of 0.3
$M_{\odot}$ \citep{Harrington1990}.  As discussed in
$\S$\ref{subsubsec:astrometry}, we see the separation between the
photocenters of the AC and B components decrease during the 4 years of
Kepler observations in a manner that is consistent with the orbit of
the AC system. Due to Kepler's large plate scale of 4\arcsec/pixel,
separate light curves for components A and B cannot be generated by
aperture photometry. That limitation motivated this work, which aims
to generate separate light curves for components A and B from the
pixel-level data.

This paper is the third in a series studying flares with Kepler. Paper
1 examined the stellar activity of 5 early-to-mid type M dwarf
systems, including GJ 1245. Paper 1 reported the rotation periods for
components A and B of $0.2632\pm0.0001$ and $0.709\pm0.001$ days,
respectively, based on light curve modulations due to starspots. We
confirm those periods within the uncertainties, which are likely due
to the effects of differential rotation discussed in
$\S$\ref{sec:spots}. Paper 1 also reported a flare sample for the AB
system, based on their combined light curve. \citet{Davenport2014},
hereafter referred to as Paper 2, focused on the active M dwarf GJ
1243, with a detailed analysis of the temporal morphology of its
flares based on a sample of over 6,100 flare events.  In this paper,
we analyze the flare properties and starspot evolution of the two
stars individually based on their separated light curves.

Containing two nearly identical and fully convective M dwarfs, the GJ
1245 system provides a unique test case to break some of the
degeneracies in stellar age-rotation-activity models. Given that stars
A and B are coeval and of nearly equal mass, but have rotation periods
that differ by almost a factor of 3, we aim to answer several simple
yet fundamental questions. Namely, which star flares more often, and
how do the energy distributions of their flares differ? We introduce a
new metric, $L_{fl}/L_{bol}$, to compare the energy emitted in flares
relative to the bolometric luminosity, and discuss the caveats of this
metric in the context of GJ 1245 A and B. As both stars exhibit
periodic brightness variations due to starspots, we also aim to
determine if their starspot properties differ, and look for evidence
of differential rotation. Here we are interested in the bulk activity
properties of the two stars, and their dependence on rotation rate.

The outline of the paper is as follows. In section $\S$\ref{sec:obs}
we describe the Kepler data, demonstrate that they contain a clear
signal from both components A and B, and describe the process used to
generate separate light curves for each component. In
$\S$\ref{sec:spots} we compare the nature and evolution of their
starspots, and in $\S$\ref{sec:flares} we identify and compare the
flares on each component. We conclude in $\S$\ref{sec:discuss} by
comparing these results to those for GJ 1243, and discuss the results
in the broader context of stellar age-rotation-activity models.


\section{Observations and Analysis}
\label{sec:obs}
The analysis presented here involves Kepler target pixel files, which
contain the raw data transmitted from the spacecraft. A full
description of Kepler data processing is given in \citet{Fanelli2011},
but we give a brief overview. The Kepler detector consists of 42 CCDs,
each of which is divided in half to create 84 output channels. Due to
onboard storage and transmission limitations, Kepler was unable to
transmit the full image of its field of view with every
exposure. Instead, only the pixels immediately surrounding targets,
referred to as target pixel masks, or ``postage stamps'', were
transmitted. A target pixel file contains all of the images of a mask
taken during an observing quarter. For each mask, an aperture around
the target was chosen.  The pixels within this aperture were used in
the Kepler Science Operations Center processing to produce the
calibrated, detrended Pre-search Data Conditioning - Simple Aperture
Photometry (PDC-SAP) light curve \citep{Smith2012}. The PDC-SAP light
curves were used for the bulk of the Kepler exoplanet investigations,
as well as the stellar activity analyses in Papers 1 and 2. In this
section we present our justification for performing our own reduction
using PyKE pixel response function (PRF) fitting models, as well as
the validation of those models.

\subsection{Kepler Target Pixel Files}
\label{subsec:pixelfile}

The GJ 1245 system was observed with two different pixel
masks. Component A (KIC 008451868) was targeted in long cadence mode
(30 minute sampling) during quarters 1 -- 17 under Guest Observer
programs 10000 and 20028. Component B (KIC 008451881) was targeted in
long cadence mode during quarters 0 -- 17, and in short cadence mode
(1 minute sampling) during quarters 8, 10, and 11 under Guest Observer
programs 20016, 20028, 20031, and 30002. Each quarter corresponds to
approximately three months of observations, with the exception of
quarters 0, 1, and 17, which are shorter. While observed as separate
objects, the target pixel files are similar and contain both
components within the masks. However, the PDC-SAP light curves are
very different because in each case the apertures were chosen to
minimize the flux from the other component. The analysis in Paper 1
used the B component data, as the short cadence observations were
taken with that mask.

The PDC-SAP light curves exhibit flares and periodic modulation due to
starspots. As reported in Figure 4 of Paper 1, a periodogram of the
light curve from the B component mask reveals two strong signals at
0.26 and 0.71 days corresponding to the rotation periods of components
A and B, respectively. The light curve from the A component pixel mask
does not contain any significant signal at the rotation period of the
B component, likely because the aperture and data reduction removed
most of the B component flux. The PDC-SAP light curve for the B
component presents a challenge, as it was taken in the short cadence
mode most sensitive to flares, but contains significant signal from
star A. Using these data, it is impossible to determine which
component is flaring, and to compare the starspot evolution of each
component individually.

This limitation motivated us to examine the target pixel files with
the hope of generating separate, uncontaminated light curves for each
component. We focused our analysis on the B component pixel mask data,
as it included short cadence observations that are necessary to detect
all but the largest energy flares. The data are stored as FITS files
containing the observation time, the raw counts in each pixel, and the
calibrated flux in each pixel. While this calibration includes
corrections such as bias subtraction and flat fielding, it does not
remove systematic instrumental trends, unlike the PDC-SAP
processing. The target pixel files also contain information such as
the aperture boundaries, the World Coordinate System (WCS)
transformations, and the instrument configuration. In the case of GJ
1245, the size of the mask region ranged in size from $7 \times 8$
pixels to $13 \times 11$ pixels. The Kepler detector has a large plate
scale of 4\arcsec/pixel, undersampling its PSF and producing images
that can initially be challenging to interpret.  Given the relatively
small number of pixels involved, we found it most effective to plot
the fluxes contained in each pixel as individual light curves.

\begin{figure*}[ht]
\centering
\includegraphics[width=7in, trim = 0.25cm 2cm 0.25cm 1.0cm, clip ]{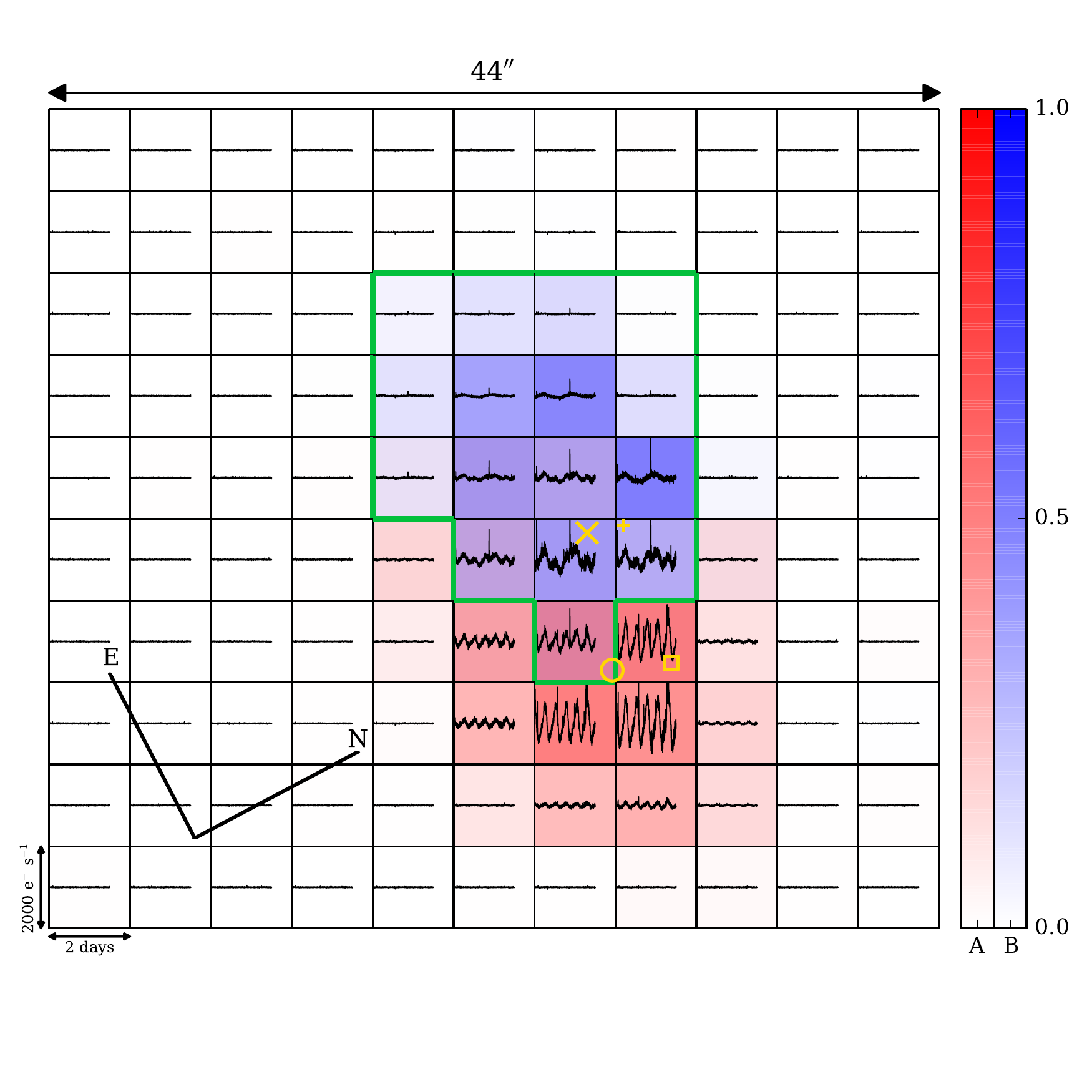}
\caption{The $11 \times 10$ grid represents the spatial extent of the
  target pixel mask. The field of view is shown by the 44{\arcsec}
  arrow at the top, and the on-sky orientation of the mask is shown by
  the arrows labeled ``E'' and ``N''. Within each cell, the
  pixel-level light curve is plotted. The y-axis range in each pixel
  is 2000 $\mathrm{e^{-}\;{s^{-1}}}$, and the timespan is 2 days, as
  denoted in the lower left corner. The color of each pixel
  corresponds to the strength of the starspot signals in each pixel,
  indicated by the color bars on the right. The locations of the PRF
  model sources for the A and B components are shown as a yellow
  circle and X, respectively. The expected positions of the stars
  based on their R.A. and decl. are plotted as a yellow open square
  and plus symbol, respectively.}
\label{fig:pixel}
\end{figure*}

An example plot of 1.5 days of Quarter 8 short cadence data is shown
in Figure \ref{fig:pixel}. The plot contains spatial, temporal, and
frequency information. The $11 \times 10$ grid represents the spatial
extent of the pixel mask, with each cell corresponding to one
pixel. The field of view is shown by the arrow labeled 44{\arcsec} at
the top of the plot. Arrows labeled ``N'' and ``E'' in the bottom left
corner show the on-sky orientation of the mask. Within each pixel,
flux is on the y-axis with a range of 2,000
$\mathrm{e^{-}\;{s^{-1}}}$, and time is on the x-axis with a range of
2.0 days, as noted in the lower left corner. Only 1.5 days of data are
shown for visual clarity. For reference, the boundary of the PDC-SAP
aperture is outlined in green. As discussed in $\S$\ref{subsec:pyke},
the locations of the PRF model sources for the A and B components are
shown as a yellow circle and X, respectively. The expected positions
of the stars based on their R.A. and decl. are plotted as a yellow
open square and plus symbol, respectively. The R.A. and decl. are
obtained from the targets' 2MASS coordinates \citep{Cutri2003}
precessed to epoch J2000.0, taking into account the proper motions in
\citet{Harrington1990}. The resulting coordinates were converted to
pixel locations using the WCS transformations contained in the target
pixel files.

The pixel-level data contain two clear periodic signals corresponding
to the two rotation periods. Pixels have been colored based on the
strength of the signal from each component, i.e., the power of the
peaks in the periodogram for that pixel. Red pixels contain a 0.26 day
signal from the A component, while blue pixels contain a lower
amplitude 0.71 day signal from the B component. Purple pixels in the
center contain signal from both components, as evidenced by the beat
pattern. White pixels are sky pixels and do not contain a significant
signal from either component. During this time period there is a flare
on the B component pixels that does not appear on the A component
pixels. This plot demonstrates both the wealth of information
contained in the target pixel files, and the feasibility of generating
separate light curves to recover individual information on starspot
modulation and flares for each component.

While the two stellar components are clearly evident in the
pixel-level data, they are not separated enough in the images to
generate uncontaminated light curves via aperture photometry. We note
that the aperture used for the B component PDC-SAP reduction, outlined
in green in Figure \ref{fig:pixel}, largely excludes pixels of the A
component. This was done consistently for all quarters across the
entire observation period, thus eliminating a large fraction of the
total flux from the system. This explains the observation made in
Paper 1 that the PDC-SAP light curve was unexpectedly noisy given the
total brightness of the GJ 1245 system. However, as the B component
PDC-SAP light curve still contains a significant signal from the A
component, it is of limited utility in studying the flare and starspot
properties of the B component individually.

\subsection{PRF Model Light Curves with PyKE}
\label{subsec:pyke}

To generate separate light curves for each component, we used the
\textit{kepprf} and \textit{kepprfphot} routines in the PyKE software
package \citep{Still2012}. Full documentation of PyKE is available on
the Kepler Guest Observer
website\footnote{\textit{keplergo.arc.nasa.gov/PyKE.shtml}}. The
\textit{kepprf} routine fits one or more sources to a target pixel
image, using a PRF model derived during spacecraft commissioning. The
\textit{kepprfphot} routine performs the same functions as
\textit{kepprf}, but generates light curves by fitting the
observations within a given time window, or an entire quarter. Here we
make a distinction between the point spread function (PSF), which is
how light falls onto the detector, and the PRF, which is how the
detector sees the PSF. The PRF can differ from the PSF due to pointing
jitter during an exposure and systematics within the detector.
 
The Kepler PRF model was derived during spacecraft commissioning by
observing approximately 19,000 calibration stars in a dither pattern
\citep{Bryson2010}. This dithering allowed the PRF to be sampled at
the sub-pixel level. The PRF model was then computed as a polynomial
fit to the dithered observations. The publicly available PRF model
used by PyKE is in the form of a lookup table. For each CCD output
channel, there are five PRF models, one defined at each corner and one
in the center. The PyKE routines linearly interpolate between the five
PRF models to generate a single model used for the fit. The model is
defined on a $50 \times 50$ grid within each pixel. Given three
user-specified free parameters for the sources: flux and the column
and row positions on the detector, the routines use the PRF model to
compute the total flux within the mask that would result from sources
with the given locations and fluxes. They then find the parameters
that minimize the residual flux across the pixel mask.

As the locations of the components on the detector vary from quarter
to quarter due to spacecraft roll and pointing changes, we first ran
\textit{kepprf} to determine their initial locations at the start of
each quarter. Knowing the on-sky separation of the two components, and
using the rotation period information seen in Figure \ref{fig:pixel},
we were able to make a reasonable guess of the source locations. As
stated in the PyKE documentation, the model convergence is not very
dependent on the initial guesses for location, as long as they are
within one pixel of the true position. Because the components are of
roughly equal luminosity, we set the initial fluxes to
equal. Convergence of the model is also not very dependent on the
initial flux values.

\begin{figure}[ht]
\centering
\includegraphics[width=3.45in, trim = 0.25cm 0.25cm 1cm 0cm, clip ]{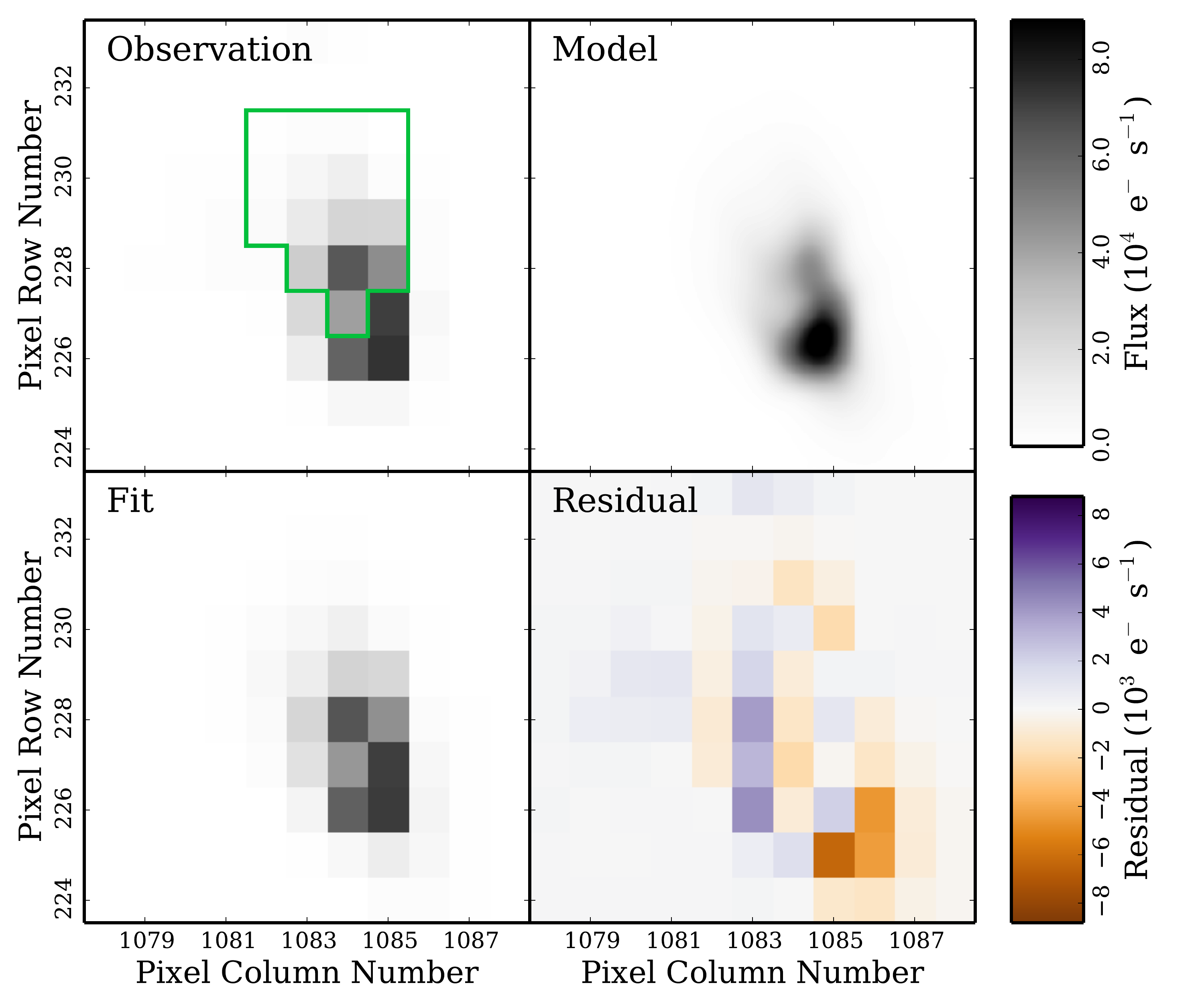}
\caption{The top left panel corresponds to a single observation of the
  target pixel mask, with the flux in each pixel indicated by the
  greyscale colorbar. The green border demarcates the PDC-SAP
  aperture. The PyKE PRF model is in the top right panel, which is
  summed within each pixel to produce the fit in the lower left
  panel. The lower right panel shows the residual between the
  observation and fit. Note that the residual color bar has both
  negative and positive values, and has a factor of 10 smaller range
  than the other panels.}
\label{fig:kepprf}
\end{figure}

A typical output of the \textit{kepprf} routine is shown in Figure
\ref{fig:kepprf}. The top left panel shows an image from the same
quarter and pixel mask as Figure \ref{fig:pixel}. Unlike Figure
\ref{fig:pixel}, the greyscale color bar signifies the flux for a
single exposure. The top right panel shows the PRF model with sources
at the A and B component locations. The locations of the A and B
components in the model are shown as the yellow circle and X,
respectively, in Figure \ref{fig:pixel}. The model flux is defined on
a $50 \times 50$ sub-pixel grid, which must be summed within each
pixel to generate the pixel-level fluxes labeled ``Fit'' in the lower
left panel. The residuals between the observation and the fit are
shown the lower right panel. Note that the color bar for this residual
panel contains both negative and positive values, and has a factor of
10 smaller range in order to show the residuals in greater
contrast. We analyze the residuals and validate the PRF model in
$\S$\ref{subsubsec:fluxratios}.

In our reduction, we chose to include exposures flagged with quality
issues, as described in the Kepler Archive Manual
\citep{Thompson2014}. Generally representing $\lesssim$ 10\% of the
data in a given quarter, some of these flagged observations are not in
fact exposures, but instead placeholders for when the spacecraft was
in safe mode, and thus contain no data. In this case
\textit{kepprfphot} does not attempt a fit. Some exposures were
flagged as containing a cosmic ray. If a genuine cosmic ray were
detected, it would be limited to a single exposure and therefore not
identified as a flare by the procedure described in
$\S$\ref{sec:flares}. It is also possible that a genuine flare event
could be mistaken for an anomaly and removed in the calibration of the
target pixel files, but the data flags indicate this is not the case
for the short cadence data used for the flare analysis in
$\S$\ref{sec:flares}.

The majority of the remaining flagged exposures were taken during
events that have the potential to degrade the photometric precision,
such as thermal equilibration after a spacecraft Earth pointing, or
scattered light falling onto the mask. These flagged exposures were
generally included in the PDC-SAP reduction, and we chose to include
them in our reduction as well. The model fits to the flagged exposures
are consistent with the unflagged exposures, although at times they
appear noisier. We concluded that the risk of degraded photometric
precision is outweighed by the benefit of increased time sampling when
searching for flares. The effect that this noise source (and others)
have on our ability to detect low amplitude flares is discussed in
$\S$\ref{sec:flares}.

We ran \textit{kepprfphot} to generate separate light curves for all
18 quarters of long cadence data and all 3 quarters of short cadence
data taken with the B component pixel mask. The calibrated fluxes in
the target pixel files are already background subtracted, so as
recommended in the documentation, we did not include a background
source. Nor did we include parameters for pixel scale variation and
focus rotation, in keeping with the documentation's
recommendation. Conservatively, we set the convergence tolerances for
the residual minimization to $10^{-7}$. Smaller values correspond to a
smaller error tolerance. We saw no change in the model output below
$10^{-6}$, so further decreasing the tolerance would not have changed
the results. The separated light curves produced by
\textit{kepprfphot} represent the source fluxes from the best-fit
models to each exposure.  With the separated light curves, we are able
to analyze the flare and starspot properties of each star
individually. However before proceeding to do so, we next validate our
model light curves to ensure that we have correctly deconvolved the
two components.

\subsection{Model Validation}
\label{subsec:validation}

To validate our PRF models, we compare our results to several
well-constrained astrophysical properties of the GJ 1245 system. These
include the on-sky location and angular separation of the stars, their
rotation periods, and their flux ratio. 

\subsubsection{Astrometry}
\label{subsubsec:astrometry}

The positions of the A and B components in the model for Quarter 8 are
plotted in Figure \ref{fig:pixel} as a yellow circle and X,
respectively. These positions correlate well with the strength of the
starspot signals shown by the red and blue color bars.  The expected
R.A. and decl. of the A and B components have been transformed into
detector coordinates using the WCS data contained in the pixel file
header, and are plotted in Figure \ref{fig:pixel} as a yellow plus and
unfilled square, respectively. The predicted locations agree well with
the models, and differ by less than a pixel, below the level at which
the model convergence is dependent. The small discrepancy could be due
to a combination of uncertainties in spacecraft pointing, the WCS
transformations, the 2MASS coordinates and proper motions used to
calculate the R.A. and decl., as well as uncertainties in the PRF
model.

Computing the angular separation of components A and B as function of
time over the four years of observations provides both a means to
verify our models, and a test of Kepler's astrometric
capabilities. The mean angular separation during the four years is
6.7$\pm$0{\farcs}2, consistent with the value of 6\farcs96 in
\citet{Dieterich2012}. We find that the measured separation can vary
significantly within a quarter, by up to a few tenths of an
arcsecond. These intra-quarter variations repeat on an annual cycle,
likely due to the stars being on different parts of the focal plane as
the spacecraft executed four seasonal rolls to keep its solar panels
pointed towards the Sun. This suggests that the intra-quarter
variations are due to systematic effects within the spacecraft optics
and detector, such as differential velocity aberration or intra-pixel
sensitivity variations \citep{Christiansen2013}. It is also possible
that the true PRF evolved as a function of time, and therefore
differed from the model derived during commissioning. The PyKE PRF
models can include parameters for pixel scale changes and PRF rotation
in the fit. A test reduction including these parameters did not
improve the fit and caused no change in the separation trends. The
PyKE documentation does not recommend including these parameters, and
we did not include them in our reductions.

\begin{figure}[ht]
\centering
\includegraphics[width=3.45in, trim = 1cm 1cm 1.5cm 1.5cm, clip ]{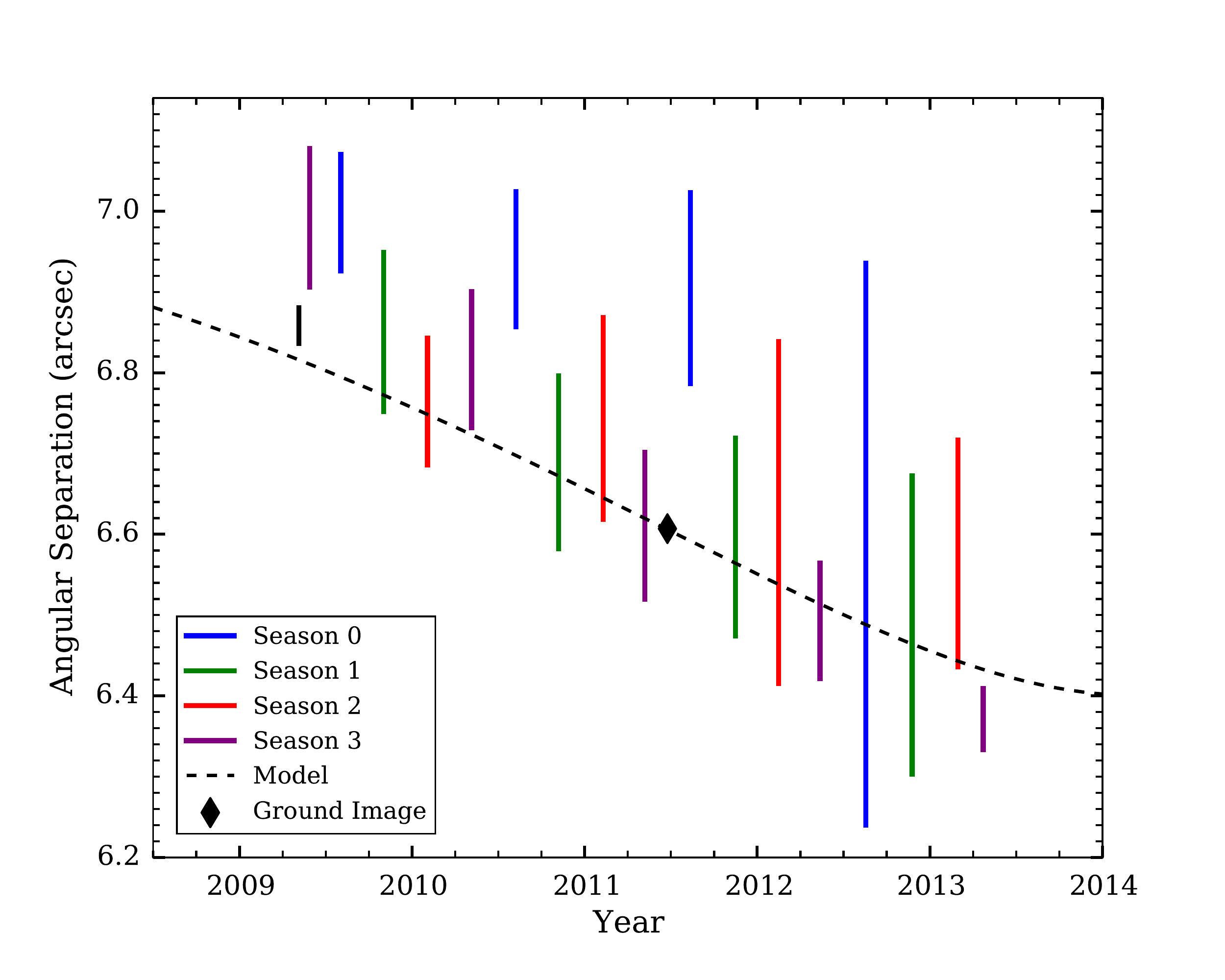}
\caption{The ranges of angular separation in each Kepler quarter are
  plotted as vertical lines. Quarters are color-coded based on the
  observing season (spacecraft orientation) during which they were
  taken. The dashed line corresponds to the expected angular
  separation based on orbital parameters derived from HST FGS
  observations. The model is constrained to pass through the
  ground-based data point, shown as a black diamond. Quarter 0 does
  not correspond to the seasonal cycle, and is plotted in
  black. Quarter 17 (rightmost purple line) is only 30 days long, and
  therefore has a smaller variation.}
\label{fig:separation}
\end{figure}

In addition the intra-quarter variations, we observe a long term
trend of decreasing separation between components A and B, indicative
of a shift in the AC photocenter caused by the unseen C component.
Plotted in Figure \ref{fig:separation} is the angular separation of
components A and B in each quarter of Kepler data, as determined by
our PRF models. The vertical lines represent the range of values
within each quarter due to the intra-quarter variations discussed
above. Data points have been color-coded based on the observing
season, i.e., spacecraft orientation, in which they were
taken. Quarter 0 was a commissioning period that does not correspond
to the seasons of the other quarters, so it is plotted in black. In
unresolved images like the Kepler observations, a binary orbit such as
GJ 1245AC would be observed as the photocenter of the two stars
orbiting their center of mass. \citet{Harrington1990} measured a
photocentric perturbation for GJ 1245AC of 0\farcs28 with a period of
15.2 years. Given its expected amplitude and period, it should be
detectable in the Kepler data.

Modeling the expected angular separation as a function of time
requires knowing both accurate orbital parameters of the AC system,
and a recent measurement of the separation between the AC and B
components.  The HST Fine Guidance Sensor (FGS) observations of GJ
1245 A and C described in \citet{Henry1999} were made as part of a
long term astrometric program, and have yielded updated orbital
parameters (Benedict et al. in prep.). The NOAO Science Archive contains a
2011 $V$ filter observation of GJ 1245 taken with the WIYN 0.9m mosaic
imager. We determined the angular separation of the AC and B
components in this image by measuring the centroids with the IRAF
\textit{imexam} tool, and then converting their detector positions to
R.A. and decl. using WCS transformations. The resulting angular
separation is 6\farcs6 at position angle (PA) 77 degrees east of
north. For comparison, the separation was 6\farcs96 at PA 83 degrees
in 1998 \citep{Dieterich2012}, and 7\farcs97 at PA 98 degrees in 1975
\citep{Harrington1990}, evidence of the several hundred year orbital
motion of the AC and B components around their center of mass.

Using the AC orbital parameters derived from the FGS observations,
along with the AC to B component separation measured in the 2011
ground-based image, we model the expected angular separation in Figure
\ref{fig:separation}. The model is constrained so that it must pass
through the ground-based data point. The slope and amplitude of the
model is consistent with the observations. The Season 0 data points
have the largest scatter, and if they are disregarded the model agrees
well with the data, given the fairly large intra-quarter
variations. We did not attempt to correct for these variations, so a
more detailed analysis and comparison to the much more precise HST
observations is beyond the scope of this paper. However, these results
serve to validate our PRF models, given their agreement with the
contemporaneous ground-based image, and that the long term trend is
consistent with an astrometric perturbation due to the C component.

\subsubsection{Rotation Periods}
\label{subsubsec:rotation}

\begin{figure*}[ht]
\centering
\includegraphics[width=7.0in, trim = 0cm 0.5cm 0cm 1.5cm, clip]{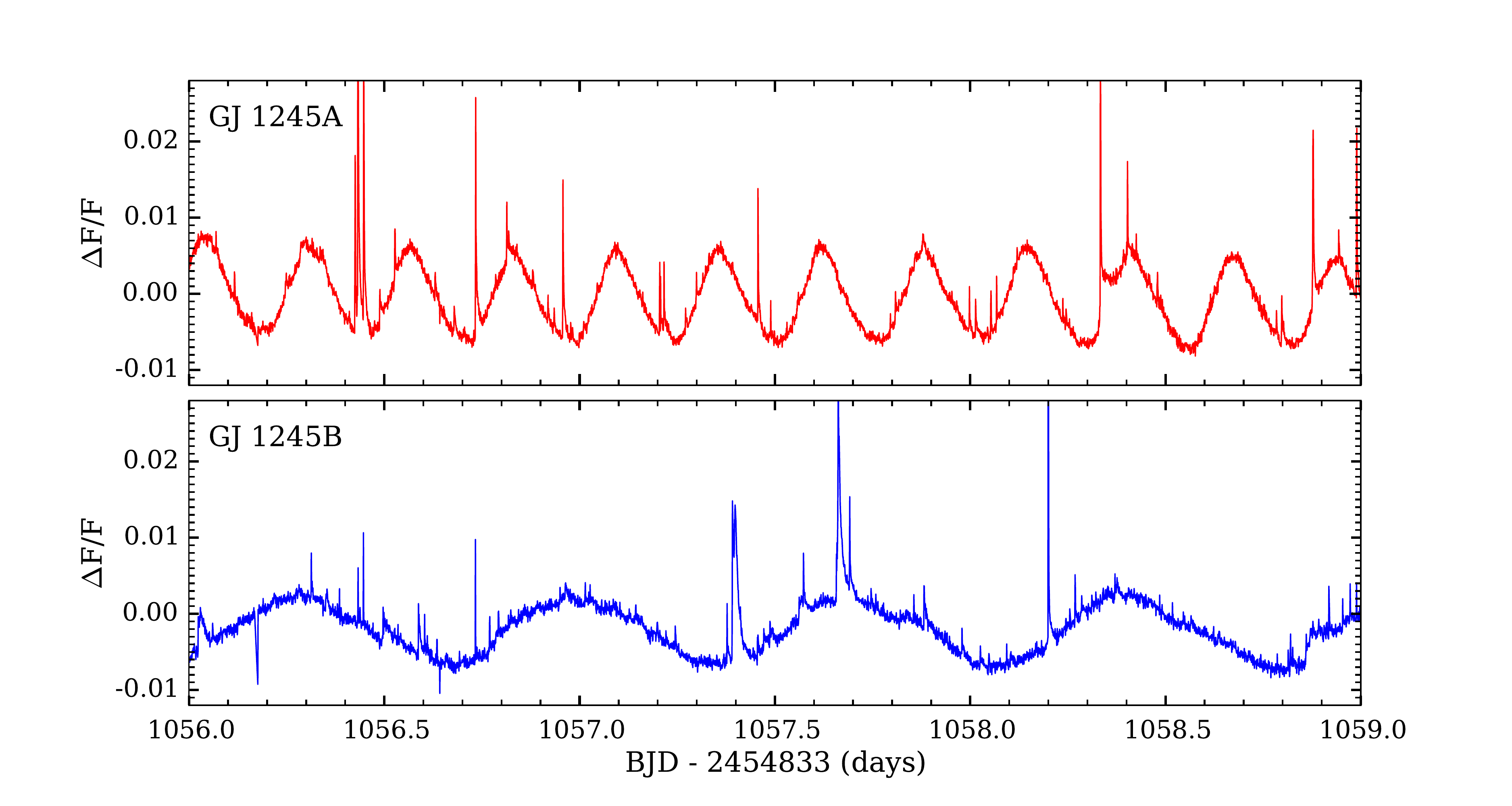}
\caption{An example of the separated, short cadence light curves
  generated by the \textit{kepprfphot} routine, in terms of relative
  flux. Nearly all of the flares shown are separate events occurring
  on only one star. The negative flux excursions are single-exposure
  errors in the models.}
\label{fig:lightcurve}
\end{figure*}

\begin{figure}[ht]
\centering
\includegraphics[width=3.0in, trim = 0.7cm 3cm 2.2cm 3.cm, clip ]{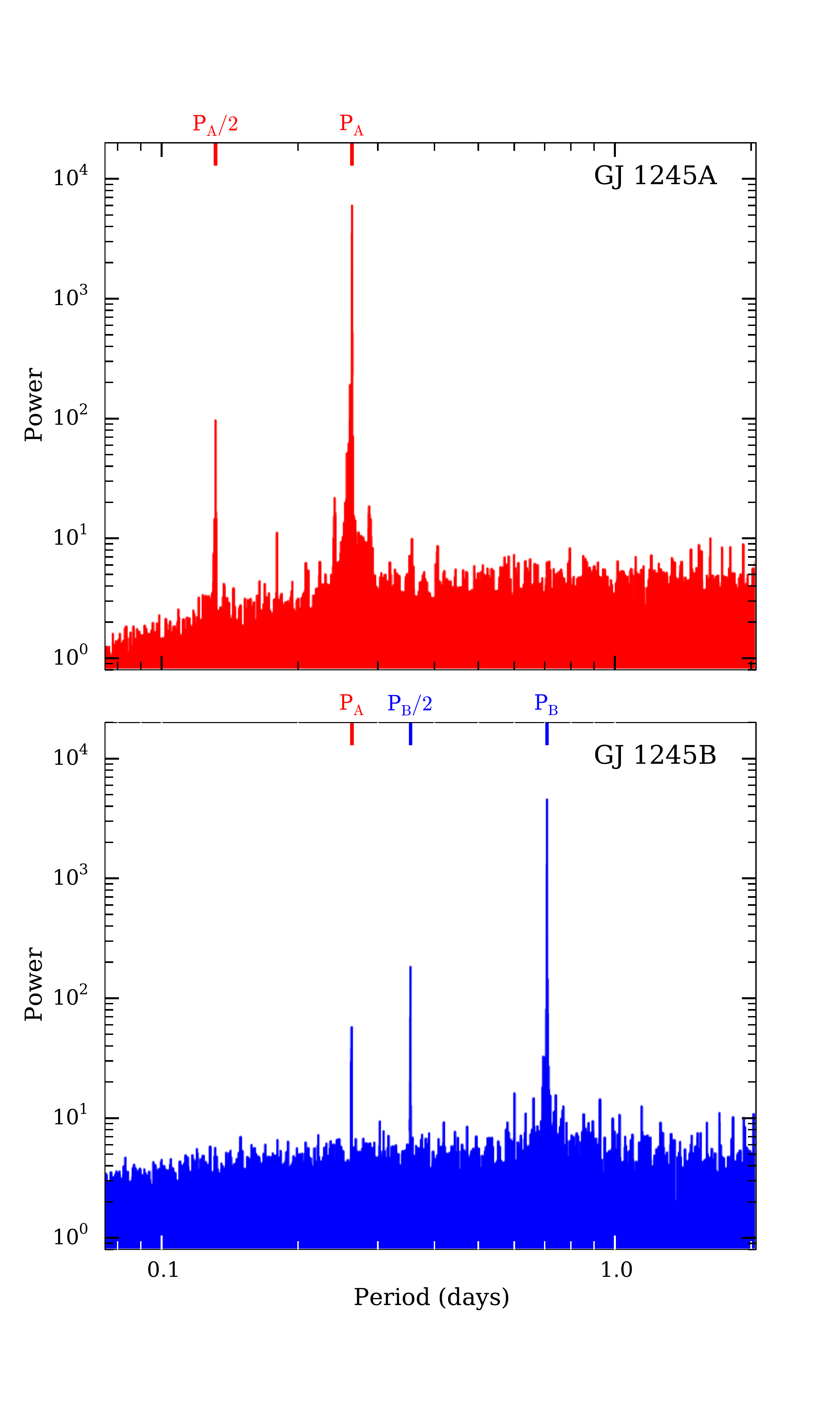}
\caption{Lomb-Scargle periodograms for the four year, long cadence,
  separated light curves. The light curves were de-trended and
  converted to units of relative flux prior to generating the
  periodograms. Peaks corresponding to the full and half rotation
  periods of stars A and B are labeled. The other small peaks in the A
  component light curve do not correspond to the rotation period of
  either star, and are likely due to noise.}
\label{fig:periodogram}
\end{figure}

An example of the short cadence separated light curves produced by
\textit{kepprfphot} is shown in Figure \ref{fig:lightcurve}. By eye,
it is easy to see the two starspot signals, as well as flares that
occur on one component but not the other.  Lomb-Scargle periodograms
of the 4 year, detrended, long cadence light curves discussed in
$\S$\ref{sec:spots} are shown in Figure \ref{fig:periodogram}. In the
A component periodogram, there is a large peak at the 0.26 day
rotation period of the A component, as well as at 1/2 of the A
period. In the B component periodogram, there is a large peak at the
0.71 day rotation period of the B component, and at 1/2 of the B
rotation period. There is also a small peak at the A component
period. These periodograms indicate that the PRF models have cleanly
deconvolved the two components with minimal cross-contamination.


\subsubsection{Flux Ratios}
\label{subsubsec:fluxratios}

The PRF models consistently converge to a flux for the A component
that is on average 2.9 times that of component B, and varies between
quarters from 2.5 to 3.3. From the published photometry, the flux
ratio of the A component to the B component is 1.84, 1.74, 1.64, and
1.57 at $B$ \citep{Dahn1976}, $V$, $R$, and $I$ \citep{Reid2004},
covering the entire Kepler bandpass. Here we neglect the small flux
contribution from the C component, as it is $\sim$94\% less bright
than the A component. In the 2011 ground-based $V$ filter image, the
flux ratio is 1.8. The A and B components in their PDC-SAP light
curves have a mean flux ratio of 2.0 over the four years of
observations. While the A component PDC-SAP light curve appears to be
uncontaminated, there is some A component contamination in the B
component PDC-SAP light curve. Removing this contamination would raise
the flux ratio somewhat higher than 2.0, in closer agreement with our
models, but further from the ground-based data. The discrepancy
between the PRF model and the ground-based photometry does not affect
our results, other than to offset the flare energy distributions (see
$\S$\ref{sec:flares}), but we discuss it here for completeness.

If the discrepant flux ratio were due to a systematic misassignment of
flux by the model, it should appear in the model residuals. The
residuals in the lower right panel of Figure \ref{fig:kepprf} are
typical of those for the model fits over the four years of data. Note
that the residuals can have both negative and positive values, as the
model can overpredict or underpredict the flux in a pixel. While the
residuals for an individual pixel are as large as a few thousand e$^-$
s$^{-1}$, this represents $\lesssim 1\%$ of the total flux in the
pixel mask of approximately 500,000 e$^-$ s$^{-1}$.

Furthermore, the residuals show no spatial correlation, in the sense
that the model does not systematically assign more flux to one
component at the expense of the other.  We verified this for each
quarter of data by examining the 15 nearest pixels to the A and B
component source locations. If a pixel was close to both components,
it was assigned to the nearest component. For these two regions of
pixels that ``belong'' to each component, we calculated the residuals
between the observation and the model. The sum of the residuals across
each region was relatively small, between a few hundred to a few
thousand e$^-$ s$^{-1}$. This is not large enough to explain the
unexpectedly large flux ratio consistently obtained by the model
fit. It appears that the PRF model is accurately reproducing the true
PRF, and that the PRF itself is causing the B component to appear
fainter relative to the A component than it is in ground-based images.

Potential sources of error in the PRF model listed in
\citet{Bryson2010} are changes in focus since the commissioning
observations, blends or variability in the stars used to compute the
model, CCD non-uniformities, and the PRF dependence on star color,
which was not modeled. The dithering observations of a finite number
of calibration stars did not sample every pixel on the detector, so
the PRF model must interpolate to be defined over the entire detector,
limiting its accuracy. These and other factors may contribute to the
residuals in the models, as well as the intra-quarter variations in
angular separation.

While the residuals within a given pixel can be relatively large, the
sum of the residuals across the mask remains small, of the order a few
hundred e$^-$ s$^{-1}$.  This means that the model is recovering all
of the flux in the mask, although in some cases it may be assigning
some flux to the wrong component. Despite the discrepant flux ratios
we obtain, the latter scenario appears unlikely, given the lack of
starspot signal contamination in the A and B component light
curves. While it is possible that our residuals and the flux ratio
could be improved by additional modeling, the PyKE routines and the
PRF model were developed by the Kepler science team, and we did not
endeavor to augment them. Based on the overall strong agreement with
the astrophysical constraints outlined above, we believe our light
curves represent the best-fit models, and can reliably be used for the
analysis that follows.

 
\section{Starspot Evolution}
\label{sec:spots}
We used the separated, long cadence light curves to analyze the
starspot evolution of each star, as the 30 minute cadence sufficiently
samples the 8 and 17 hour period starspot modulations. In this section
we describe how long-term trends were removed from the light curves,
and examine the evolution of the starspot features.

\subsection{Light Curve De-trending}
\label{subsec:detrend}

The separated light curves exhibit smooth, long-term trends that are
typical of uncalibrated Kepler data.  M dwarfs have been observed to
exhibit long-term $VRI$ photometric variability of up to 5\% on
multi-year timescales \citep{Hosey2014}. Such variations could in
principle be detectable in our data. Unfortunately, because Kepler was
not designed for absolute photometry, we are unable to determine
whether the observed long term trends are physical or due to
instrument systematics. Because our starspot analysis is concerned
with short-term changes in the relative brightness of each star, we
simply removed these long term trends and normalized the light curve
into units of relative flux as described below.

We first smoothed the light curve using the one-dimensional Gaussian
filter in the Python SciPy package, in order to trace the low
frequency trends in the light curves without affecting the higher
frequency starspot signals. A Gaussian filter functions as a low pass
filter, and has a Gaussian frequency response function. The standard
deviation, $\sigma$, of the kernel determines the cutoff frequency of
the filter. Increasing $\sigma$ decreases the cutoff frequency. In
addition to long-term trends, the light curves also have some
discontinuities that occur at gaps in data. We addressed this by
identifying all data gaps longer than 0.5 days, and smoothing each
section of light curve separately. There remained a few
discontinuities that did not occur at data gaps, which we also
analyzed individually.

So that flares did not skew the de-trending, we performed a initial
smoothing with a kernel size of 10$\sigma$, and then rejected all
points on the original light curve that were more than two standard
deviations away from the smoothed light curve.  We then smoothed the
original light curve, with flares removed, using a kernel size of
40$\sigma$. We chose this kernel size because a Gaussian filtering of
evenly spaced data at a 30 minute cadence with a 40$\sigma$ kernel
completely attenuates all signals below 1 day. The resulting smoothed
light curve does not contain the starspot modulations, but traces the
long term trends in the original light curve. The final, de-trended
light curve was produced by subtracting the smoothed light curve from
the original light curve, and then dividing by the median flux value
of the entire un-smoothed four year light curve. The de-trended light
curve has units of relative flux, or
\begin{equation}
\frac{\Delta F}{F} = \frac{f - f_s}{f_0}
\end{equation}
where $f$ is the flux in the original light curve, $f_s$ is the flux
in the smoothed light curve, and $f_0$ is the median value of $f$ over
the four year dataset. We stress that the purpose of this de-trending
is to trace the low frequency trends and convert the light curve to
units of relative flux. The size of the Gaussian kernel was chosen so
that the de-trending did not affect the short period starspot signals.

\subsection{Differential Rotation}
\label{subsec:diffrot}

Because the light curves represent the integrated flux from the
hemisphere of the star visible at a given time as the star rotates, we
are limited in our ability to determine the spatial distribution of
the spot regions. For instance, we cannot say whether there are a few
large spots or many small spots distributed over the star. We
therefore refer to the light curve modulations as dark and light
``features'', with the minimum in the modulation corresponding to the
visible hemisphere of the star that has the largest amount of spot
coverage.

\begin{figure}[ht]
\centering
\includegraphics[width=3.5in, trim = 0cm 0cm 0cm 0cm, clip ]{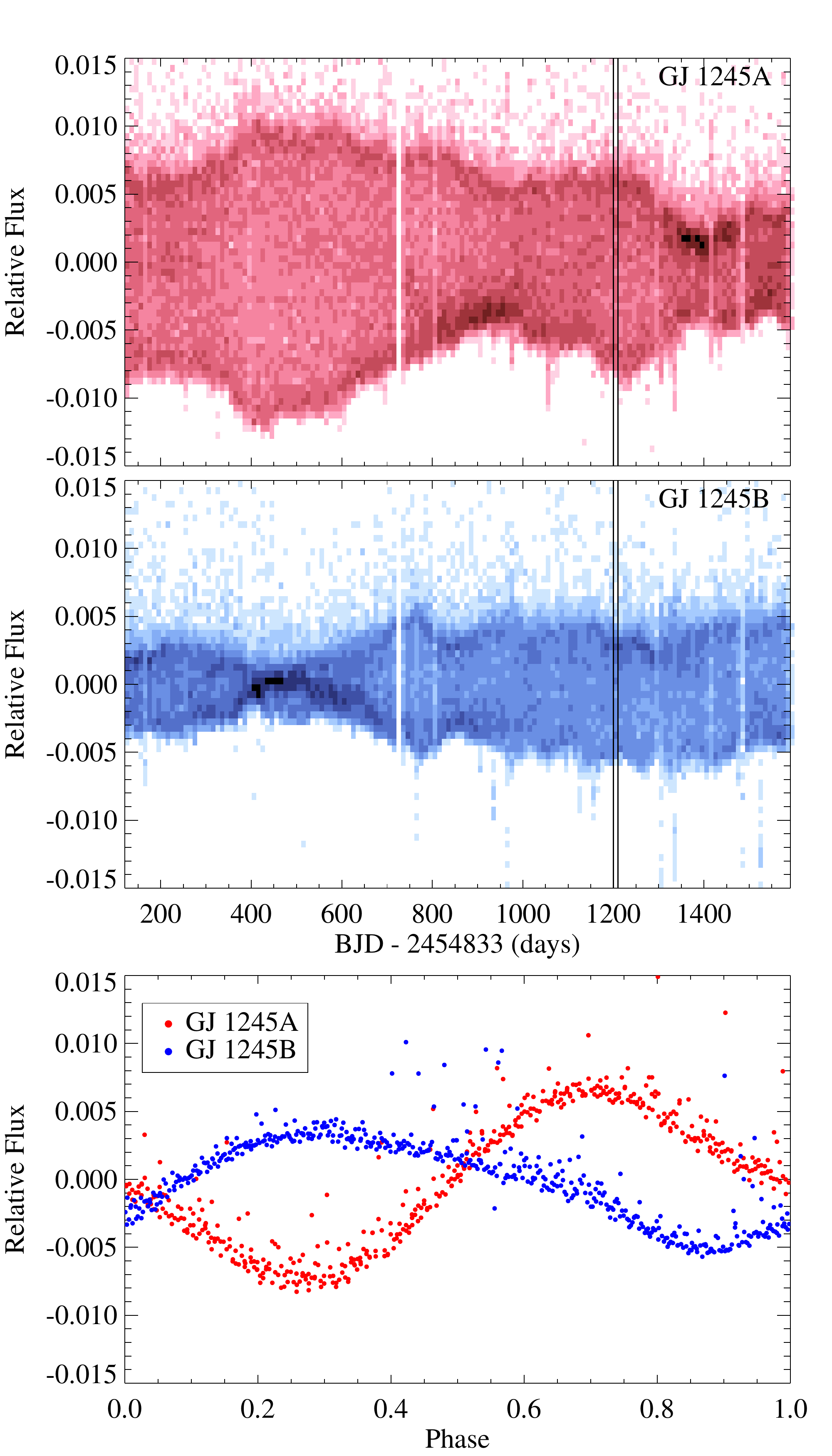}
\caption{The top two panels show the relative amplitudes of
    the starspot modulations versus time, with the light curves
    averaged in 10 day bins. In the bottom panel, phase-folded light
    curves are plotted for the bins represented by the vertical lines
    in the top two panels. The contours in the top two panels
    correspond to the density of points in the bottom panel. Note how
    the amplitude of the light curves in the bottom panel corresponds
    to the amplitude of the contours in the top two panels. Flares
  are shown as positive flux excursions, while the negative excursions
  are due to small errors in the detrending discussed in
  $\S$\ref{sec:spots}.}
\label{fig:amplitude}
\end{figure}

However, there are some basic measures of the starspot evolution that
can be obtained from the light curves. Figure \ref{fig:amplitude}
shows the relative amplitudes of the starspot modulations as a
function of time, with the light curves averaged in 10 day bins. The A
component modulation generally has a larger relative amplitude, but
the modulations on both stars show significant variations in amplitude
over time. As a check of our models, we performed the same de-trending
procedure on the long cadence PDC-SAP light curve for the A component,
and the relative amplitude evolution appears nearly identical to that
from our A component model light curve in Figure
\ref{fig:amplitude}. The changes in starspot modulation amplitude are
consistent with the spectropolarimetric results of \citet{Morin2010},
who saw changes in the large scale magnetic field on GJ 1245B during a
three year observing campaign.

\begin{figure}[ht]
\centering
\includegraphics[width=3.45in, trim = 0cm 0cm 0cm 0cm, clip ]{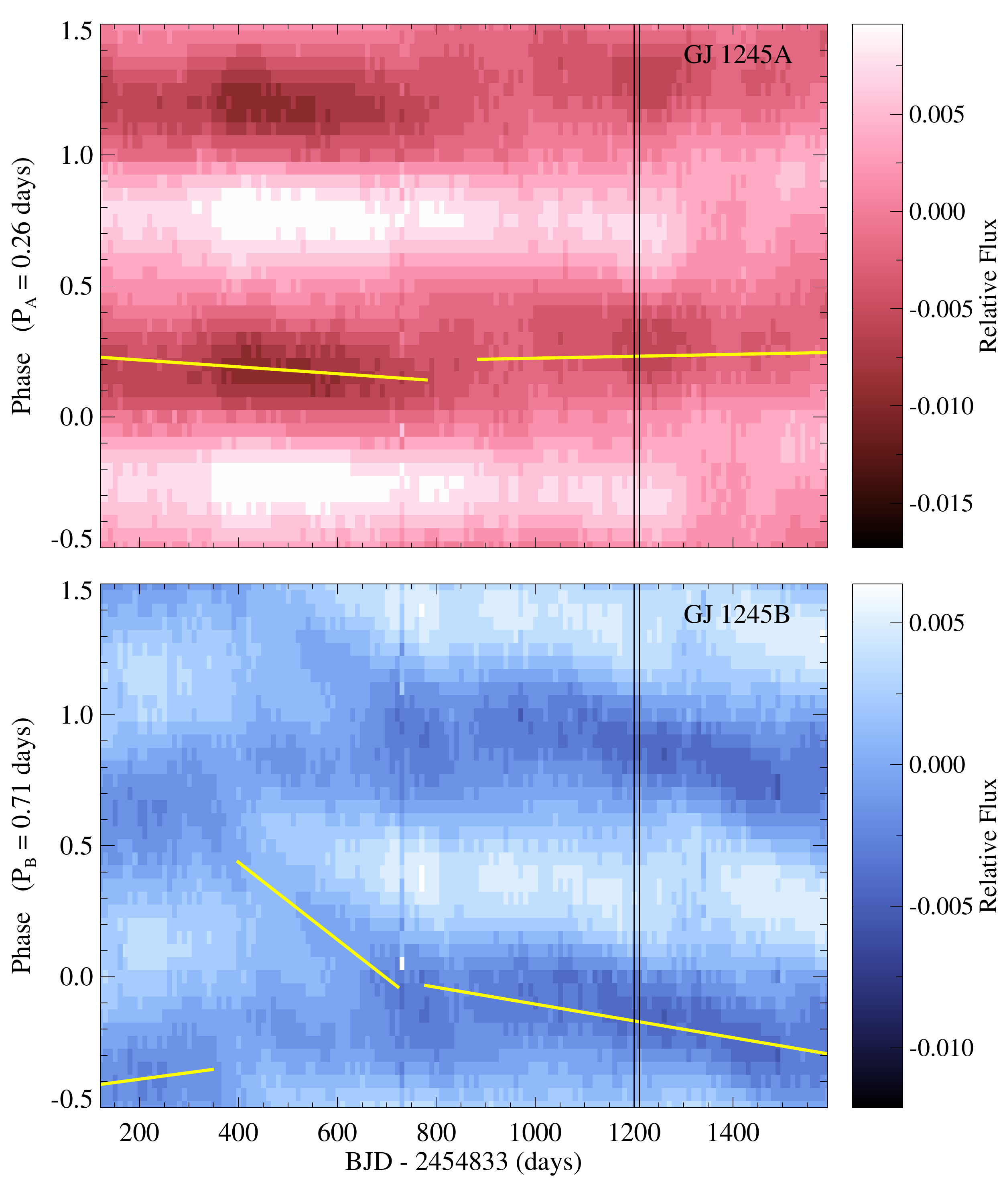}
\caption{Starspot modulation phase versus time. The light curves have
  been folded over two phases at the respective rotation period of
  each star, and averaged in 10 day bins. In this figure, the
    contours represent relative flux. For the bins represented by the
    vertical black lines, note how the darkest contours correspond to
    the minima in the phase-folded light curves in the bottom of panel
    of Figure \ref{fig:amplitude}, at phase 0.3 and 0.9 for components
    A and B, respectively. We fit the spot features with bivariate
  Gaussians. Yellow lines represent the 2$\sigma$ values along the
  time axis of the Gaussians.}
\label{fig:evolution}
\end{figure}

Another measure of the starspot evolution is to examine if there is
any phase shift in the starspot modulation. We assume that the
modulations are due to darker spot regions rotating into and out of
view, changing the integrated flux from the visible hemisphere of the
star. Thus the starspot modulation as a function of rotation phase
gives an indication of how the spots are distributed longitudinally on
the star. In Figure \ref{fig:evolution}, the light curves have been
phase-folded at the respective rotation period of each star, and then
averaged in 10 day bins. The light curves have been folded over two
phases for visual clarity. A more detailed description of this
phase-folding procedure and its application to modeling starspot
features will be given in a forthcoming paper (Davenport et al. in
prep). We fit the starspot features with bivariate Gaussians, where
the x and y-dimensions are time and phase, and the z-dimension is
relative flux. A cut through a bivariate Gaussian along the x-y plane
creates an ellipse.  In Figure \ref{fig:evolution}, we cut through
each Gaussian at its 2$\sigma$ value, and represent the time axes of
the resulting ellipses as yellow lines. The purpose of these fits is
to guide the eye, and to enable a quantitative discussion of the spot
evolution.

These stars are remarkable for the long-lived nature of their spot
features.  On the A component, we fit two spot features, both of which
remain at nearly constant phase. This could also be interpreted as a
single, long-lived spot. On the B component, we fit three spot
features, all of which show a more rapid phase evolution than the
features on the A component. Most notably, there is a dark feature
migrating from phase $\sim$0.5 to $\sim$0.0 between days $\sim$350 and
$\sim$700. This feature coincides with the minimum in the relative
amplitude seen in Figure \ref{fig:amplitude}, indicating that the spot
coverage was temporarily more evenly distributed with longitude. As is
the case for the entire light curve, there is almost no signal present
from the A component in this section of the B component light
curve. This rules out the migrating feature being caused by
contamination from the A component.

The phase evolution of these features can be explained by differential
rotation. For example, a spot near the pole and a spot near the
equator would appear to shift in phase relative to each other if there
is a variation in rotation rate with latitude. The features in
  Figure \ref{fig:evolution} could potentially be due instead to the
  meridional flow of spots, or the emergence and disappearance of
  spots. However for component B, the former effect is too slow to
  explain the rapid phase evolution, and the latter is unlikely to
  result in the coherent phase evolution that is observed. The rate of
change of phase with time, $\Delta\Omega$, reported in Table
\ref{tab:compare}, gives a lower limit on the difference in rotation
rate between the equator and the pole. Differential rotation also
affects the measured period of the starspot modulation if spots are
present at different latitudes. This creates some uncertainty in the
rotation period determination. Analyzing each of the 9 sub-quarters
($\sim$1 month in duration) of short cadence separated light curves
individually, we find mean rotation periods of 0.2632 $\pm$ 0.0001 and
0.709 $\pm$ 0.001 days for the A and B components, respectively. While
the exact period determined depends on the subset of data analyzed and
its duration, we confirm the rotation periods reported in Paper 1
within the uncertainties quoted above.

\section{Flares}
The 1 minute sampling of the short cadence data is most sensitive to
flares. Although some flares are evident in the long cadence data, we
have found that the 30 minute sampling makes it difficult to directly
compare the measured energies to the short cadence data. Therefore we
limit our flare analysis to the separated short cadence light curves,
which comprise 3 quarters of Kepler monitoring. In this section, we
describe how flares were identified in the light curves, and how the
flare samples were prepared. We next discuss the effects of non-linear
CCD response for high energy flares. Finally, we determine flare rates
for each star, characterize their power law distributions, and
determine the fraction of the stars' total energy emitted in flares.

\label{sec:flares}

\subsection{Flare Identification}
\label{subsec:flareid}

We identified flares using the automated selection procedure
described in $\S$2.1 of Paper 2. Briefly, the light curves were first
iteratively smoothed to remove the periodic starspot modulations. This
step is necessary because many flares have amplitudes smaller than the
starspot modulation, and therefore would not be identified by a simple
threshold search of the unmodified light curve. We note that this is a
different smoothing procedure than the one described in
$\S$\ref{sec:spots}. Here the function is to remove the starspot
signal but preserve the flares. Flare candidates were identified as
two or more consecutive observations with positive flux excursions
more than 2.5 times the standard deviation of the smoothed light
curve. The light curves with the tagged flare candidates were visually
inspected to ensure that the selection procedure did not mistakenly
identify data gaps or discontinuities as flares. Instances where this
occurred were removed from the sample.

Due to noise in the light curves, there is a minimum flare energy
below which we cannot reliably identify flares. Here we determine the
energy of a flare in terms of its equivalent duration, i.e., the area
under the flare light curve, measured in relative flux units. The
calculation of equivalent duration is discussed in greater detail in
\citet{Hunt-Walker2012}. Equivalent duration has units of time, but is
not to be confused with the duration of time over which the flare
occurred. Multiplying the equivalent duration by the quiescent
luminosity of the star gives the flare energy.  

For a flare of a given equivalent duration, $\mathcal{E}$, and duration,
$\tau$, in time, we define the signal-to-noise (S/N) of the flare as
\begin{equation}
\mathrm{S/N} = \frac{\mathcal{E}}{\sqrt{\mathcal{E} + \sigma\tau}}
\end{equation}
where $\sigma$ is the standard deviation of the ``continuum'' light
curve around the flare. Each flare candidate was visually inspected,
in descending order of S/N. For both stars, spurious flare events
began to contaminate the sample at a S/N value of 1.4. We therefore
excluded flare candidates below this threshold. The S/N of flares
correlates with equivalent duration, and therefore energy.  Although a
S/N value of 1.4 corresponds to the same equivalent duration value of
2.3 seconds for both stars, it corresponds to a lower flare energy
limit on the B component, because a lower energy flare is easier to
detect above the quiescent flux of the intrinsically fainter B
component. Although many of the flare candidates below the threshold
are real flare events, we set the threshold conservatively high to
limit the impact of systematic effects in the model light curves.

As determined in $\S$2.3 of Paper 1, the total quiescent luminosity of
the GJ 1245 system in the Kepler bandpass is $\log{L_{Kp}}$ = 30.22
erg s$^{-1}$. The quiescent luminosity was determined using the
apparent Kepler magnitude of the GJ 1245 system listed in the Kepler
Input Catalog, the zero-point of the Kepler magnitude system, and the
trigonometric distance of the system. The individual quiescent
luminosities of components A and B can be found from the total system
luminosity if the flux ratio of the two stars in the Kepler bandpass
is known, neglecting the small flux contribution of the C component.
We adopt the flux ratio of 1.64 in the $R$ filter based on the values
in \citet{Reid2004}. Among the standard photometric filters, the $R$
filter is most representative of the Kepler wavelength response.  This
yields individual quiescent luminosities of $\log{L_{Kp}}$ = 30.01 and
29.80 erg s$^{-1}$ for the A and B components, respectively. We note
that this adopted flux ratio is significantly smaller than that of our
PRF models. Given the uncertainties associated with the Kepler PRF, we
prefer the well calibrated optical photometry. We assume that the
discrepant PRF model flux ratio is the result of the spacecraft optics
and detector, that changes in brightness due to starpots and flares
are proportional to the baseline, quiescent brightness. Thus this does
not affect the measured relative flare energies, i.e, the equivalent
durations.

Because the light curves were produced by fitting a PRF model to the
pixel-level data, it is possible that a flare event, particularly one
with a large amplitude, could appear in the pixels of both stars and
be included in both light curves despite having originated from only
one component. In this case, the flares in each light curve should
reach their peak fluxes at the same time, and have similar light curve
morphologies. Approximately 4\% of the flares in the sample for each
star had peak times that differed by less than 3 minutes from a flare
on the other star above a S/N of 1.4. The remaining 96\%
were considered as separate events. Given the 1 minute time sampling
of the short cadence data, two events occurring more than 3 minutes
apart should easily be resolved.

The 4\% of flares that overlapped by less than 3 minutes were
discarded from the samples for both stars, with the exception of a few
flares for which the equivalent duration was a factor of 10 greater on
one star than the overlapping flare on the other star. In these cases,
we were confident as to which component the flare originated from, and
kept the larger equivalent duration flare while discarding the
smaller. The overlapping flares present several issues for the flare
samples. The A component is brighter, so a flare from component A is
more likely to contaminate component B than vice versa. However, as
they only represent 4\% percent of the flares, this does not
significantly bias the sample. It is possible that some of the
discarded overlapping flares are in fact two separate events that
happened to occur on each star at nearly the same time. The likelihood
of such events is rare, and because flares occur randomly in time this
should not affect the relative occurrence rates for each star.

We have neglected the contribution of the C component to the flare
sample for the A component. While the C component may flare often,
nearly all of its flares would be undetectable in the Kepler data. For
a flare on the C component to be visible in the Kepler data, it must
be brighter than the quiescent flux of the A component. This would
require a flux enhancement of over 3 magnitudes. The flare frequency
distributions (FFDs) of \citet{Hilton2011} put an upper limit on the
occurrence of such flares on M6 -- M8 stars at approximately once per
100 hours. This is a conservative upper limit on the occurrence rate,
as the flare sample was based largely on M6 and M7 stars, so the
occurrence rate of such large flares would be lower for an M8 analog
to GJ 1245C. During the 5,491 hours of Kepler short cadence exposures
of the GJ 1245 system, we would expect to detect no more than 55
flares from the C component. However, such a flare would only remain
at its peak, detectable brightness for a short time. They would appear
as short-lived, low energy flares in the A component light curve, and
would have been discarded by the minimum signal to noise threshold
applied to the sample.

\subsection{CCD Non-linearity and Saturation}
\label{subsec:nonlinear}

Upon investigation of the target pixel files, we found that some high
energy flares caused the Kepler CCD to respond non-linearly, and in
some cases saturate. The effect on our flare samples is discussed in
$\S$\ref{subsec:flarestats}, but we first provide some details of our
analysis, as they are relevant to other investigations of impulsive
phenomena in the Kepler dataset. We examined the raw counts in the
short cadence target pixel files, focusing on the brightest pixels for
each component, as they are the most likely to respond non-linearly
during a flare. It was crucial to inspect the uncalibrated raw counts,
because the calibrated fluxes can disguise the effects of
non-linearity and saturation. During Quarters 8, 10, and 11, the
median counts in the brightest pixel for the A component were 46\%,
83\%, and 50\% of the full well depth (10093 ADU;
\citealt{VanCleve2009}). The corresponding values for the B component
were 37\%, 45\%, and 45\%. Because the stars changed locations in the
focal plane between spacecraft rolls, roughly half of the flux from
the A component was concentrated in a single pixel during Quarter
10. In Quarters 8 and 11, the flux was distributed more evenly across
several pixels, as was the flux from the B component during all three
quarters.  As the quiescent counts in these pixels are already a
significant fraction of the full well depth, a flare that temporarily
increases the counts in a pixel by more than a factor of two or three
is a cause for concern.

The classical flare temporal morphology discussed in $\S$ 2.2 of Paper
1 is characterized by a rapid rise in flux, followed by a rapid decay,
and then a slower exponential decay. A short cadence observation
represents the sum of nine 6 second exposures
\citep{VanCleve2009}. Given the rapid rise and decay of a typical
flare, a pixel could reach its full well depth in one or more these
exposures even if the mean counts for all nine exposures is less than
the full well depth. In this case the fluxes measured for the flare
would only be lower limits. Because the CCD responded non-linearly,
and because no information is retained on the individual 6 second
exposures, we are unable to quantify the extent of the energy
underestimation.

Although up to 25\% and 9\% of the flares in the sample for
components A and B, respectively, may have caused the detector to respond
non-linearly and potentially saturate, it does not appear that this
caused any significant CCD bleeding effects onto adjacent pixels, with
one notable exception. For the largest amplitude flare in the
observations, bleeding along two pixel columns is evident, and at
least 13 pixels are saturated during the brightest point in the
flare. Additional bleeding likely occurred in pixels outside of the
target pixel mask. The flare appears in the separated light curves of
both stars, and was discarded from the sample under the criteria
described previously. 

\begin{deluxetable}{cccc}
\tabletypesize{\small}
\centering
\tablewidth{0pt}
\setlength{\tabcolsep}{0.1in}
\tablecaption{Flare Statistics}

\tablehead{
\colhead{Star}&     
\colhead{\# Flares}&
\colhead{Flares}&
\colhead{Range}\\
\colhead{}&
\colhead{}&
\colhead{per}&
\colhead{$\log{E_{Kp}}$}\\
\colhead{}                 &
\colhead{}                 &
\colhead{day}                 &
\colhead{(erg)}                 }
\startdata
GJ 1245A &  683 & 3.0 & 30.38 - 32.63$^{\star}$ \\
GJ 1245B &  605 & 2.6 & 30.16 - 33.14$^{\star}$ \\
\enddata
\label{tab:flares}
\tablenotetext{$\star$}{This sample is comprised of all flares that met the
  sample criteria described in $\S$\ref{subsec:flareid}, including
  those affected by CCD non-linearity. The upper energy ranges
  are therefore lower limits.}
\end{deluxetable}

\subsection{Flare Rates and Statistics}
\label{subsec:flarestats}

\begin{figure}[ht]
\centering
\includegraphics[width=3.5in, trim = 0.5cm 0.2cm 1.7cm 1.5cm, clip ]{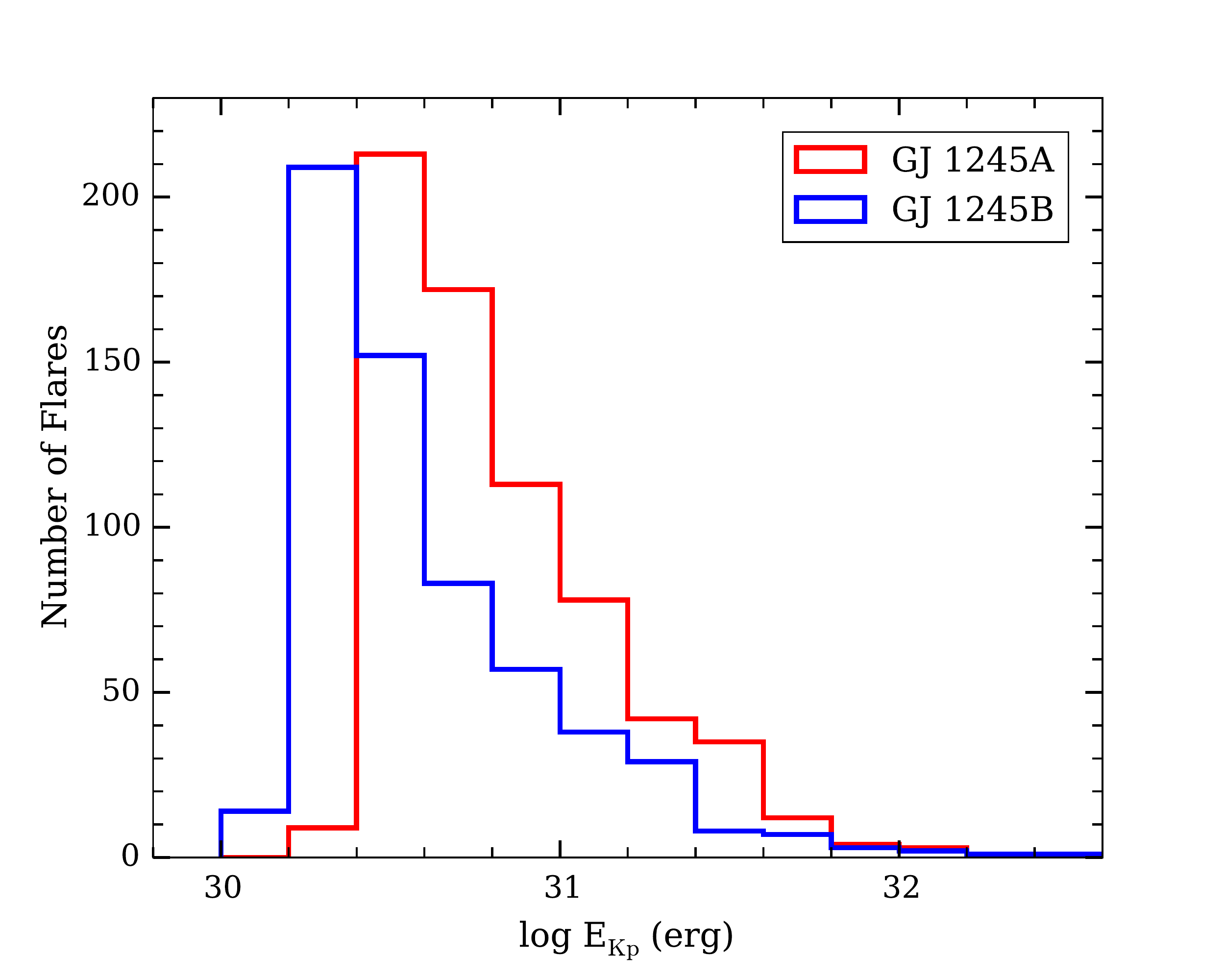}
\caption{Flare energy histograms for stars A and B. The sharp cutoffs
  at low energy are due to the minimum signal to noise threshold
  assigned to the flare sample.}
\label{fig:flarehist}
\end{figure}

The statistics of the flare samples are summarized in Table
\ref{tab:flares}, and the flare energies are plotted as a histogram in
Figure \ref{fig:flarehist}. This sample is comprised of all flares
that met the sample criteria described in $\S$\ref{subsec:flareid},
including those affected by CCD non-linearity. The sharp turnover at
lower energies is the result of the signal to noise cutoff in the
flare samples.  We note that by virtue of being on the same pixel
masks, the two stars were observed for the same amount of time, so
their flare distributions can be compared without normalization. The
energy distribution histograms are similar, but are offset in energy
due to the different quiescent luminosities of the two stars. For
flares above the S/N threshold, the average rates are also similar,
with the A component exhibiting 3.0 flares per day, compared to 2.6 on
the B component. These similarities are somewhat unexpected if
activity correlates with rotation period, given that the A component
rotates almost three times faster than B.

\begin{figure}[ht]
\centering
\includegraphics[width=3.5in, trim = 0cm 2cm 1.5cm 3.0cm, clip ]{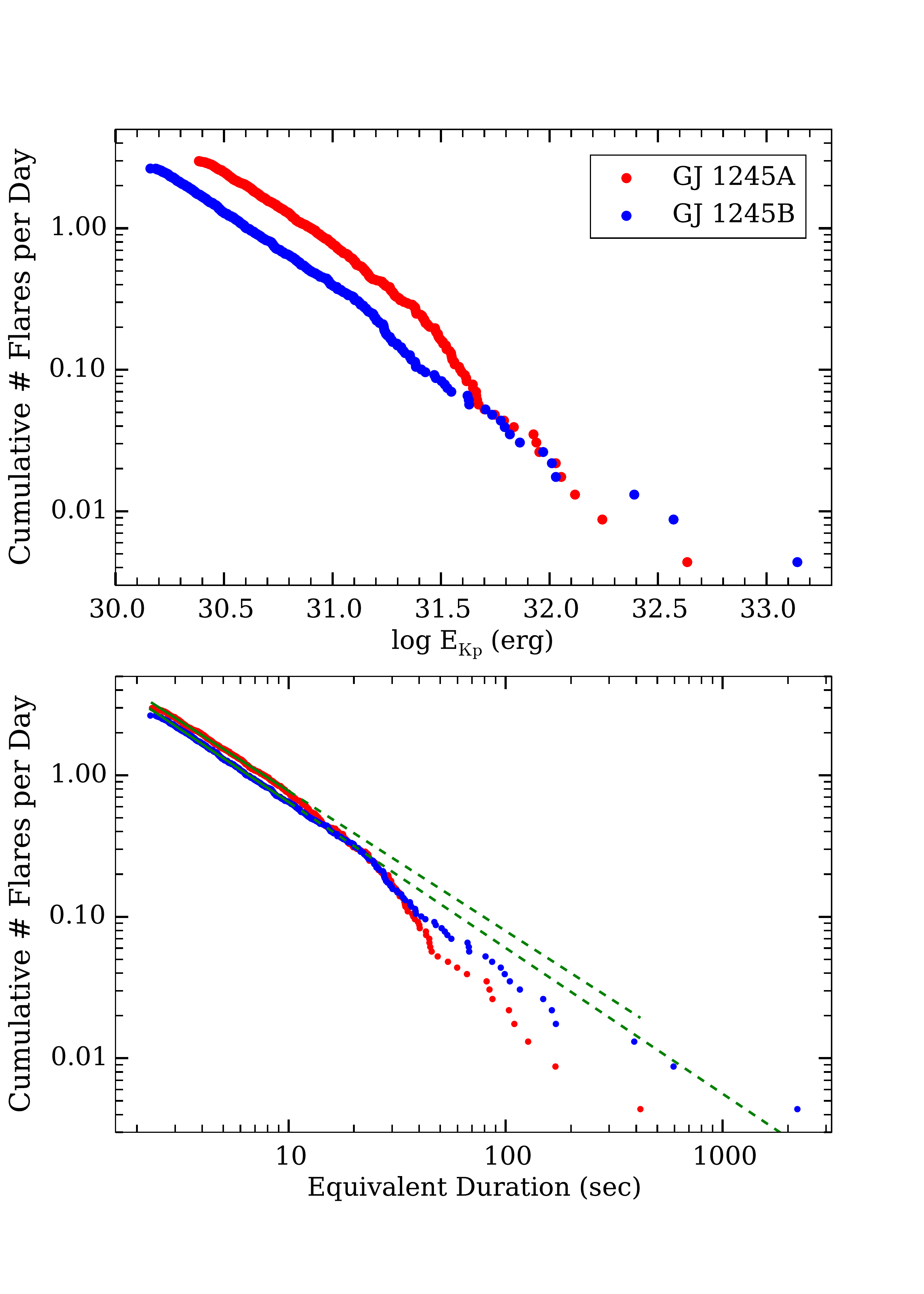}
\caption{The cumulative flare frequency distributions for stars A
  and B are plotted versus energy (top panel) and equivalent duration
  (bottom panel). For a given energy (or equivalent duration) on the
  x-axis, the cumulative number of flares per day greater than or
  equal to that energy is given on the y-axis. The power law fits
  (solid green lines) do not include flares with $\mathrm{E_{Kp}} >
  32.3$ (dashed green lines).}
\label{fig:FFD}
\end{figure}

The cumulative FFDs for components A and B are shown in Figure
\ref{fig:FFD}. The FFD gives the cumulative number of flares greater than
or equal to the given energy that occur each day. Flare frequency is
plotted versus energy on the top panel of Figure \ref{fig:FFD}, and
versus equivalent duration in the bottom panel. While it is useful to
present the FFD in physical units, this is not a fair representation
of the relative activity of the two stars. In the top panel, most of
the offset between the two FFDs is due to the quiescent luminosity
difference of the stars. In terms of equivalent duration, the FFDs lie
closer together. The equivalent duration distribution represents the
energy released in flares relative to the total energy output of the
star (see Eqn. 6 below).

Flare occurrence is typically modeled using a power law distribution
in energy of the form
\begin{equation}
N(E)dE = \beta E^{-\alpha}dE
\end{equation}
where $\beta$ is a constant. The slope of the
cumulative FFD is equal to $1 - \alpha$.  As seen in the bottom panel
of Figure \ref{fig:FFD}, the FFDs for both stars are not fit by single power
laws, shown as green lines. This is due to the non-linear response of
the CCD to high energy flares (see
$\S$\ref{subsec:nonlinear}). Because the equivalent duration (and
energy) measurements for these flares are only lower limits, we
underestimate the frequency of high energy flares, causing the FFD to
artificially steepen at high energies. The A component FFD begins to
deviate from a single power law at an equivalent duration of
approximately 10 seconds. The B component FFD is well fit by a single
power law up to an equivalent duration of approximately 25 seconds. As
expected, the deviation from a single power law is more pronounced for
the A component. Due to its greater apparent brightness, a lower
energy flare on the A component will cause its corresponding pixels to
respond non-linearly compared to the fainter B component. Due to the
non-linear response of the CCD, we are unable to determine if there is
any intrinsic change in the power law slope at higher energies.

The measured energies for the high energy flares in our sample are
compromised due to non-linear CCD response, and they are not included
in the power law fits. Some turnover is evident in the FFDs below an
equivalent duration of 2.7 seconds, due to incomplete detection of the
lowest energy flares. We therefore fit power laws to the FFDs over an
equivalent duration range of 2.7 -- 10.0 seconds for each star. We
assume that the samples are complete in this range, that these flares
did not cause the detector to respond non-linearly, and that their
energies are well constrained. The uncertainty in the number of
observed flares at a given equivalent duration was assumed to follow a
Poisson distribution. The largest contribution of uncertainty in the
equivalent duration measurements is due to fitting of the underlying
starspot modulation. This sets the baseline flux against which the
flare is measured. We found that changes in the size of the window of
quiescent light curve around the flare that is fit can change the
measured equivalent duration by up to 10\%. In most cases, the change
was less than a few percent, but we conservatively set the
uncertainties on all measured equivalent durations at 10\%.

The power law fit was performed using a Bayesian Markov chain Monte
Carlo-based algorithm \citep{Kelly2007}. The slopes with uncertainties
are reported in Table \ref{tab:compare}.  The FFDs for GJ 1245 A and B
presented here supersede the combined FFD of Paper 1, which was based
on the combined PDC-SAP light curve, and included flares from both
stars. The power law slope for the combined FFD in Paper 1 was $-1.32$
($\alpha = 2.32$), steeper than what we report here for the individual
stars as a result of including the lower limits for the high energy
flares in the power law fit.

\begin{figure}[ht]
\centering
\includegraphics[width=3.5in, trim = 0.25cm 1.75cm 1.5cm 1.75cm, clip ]{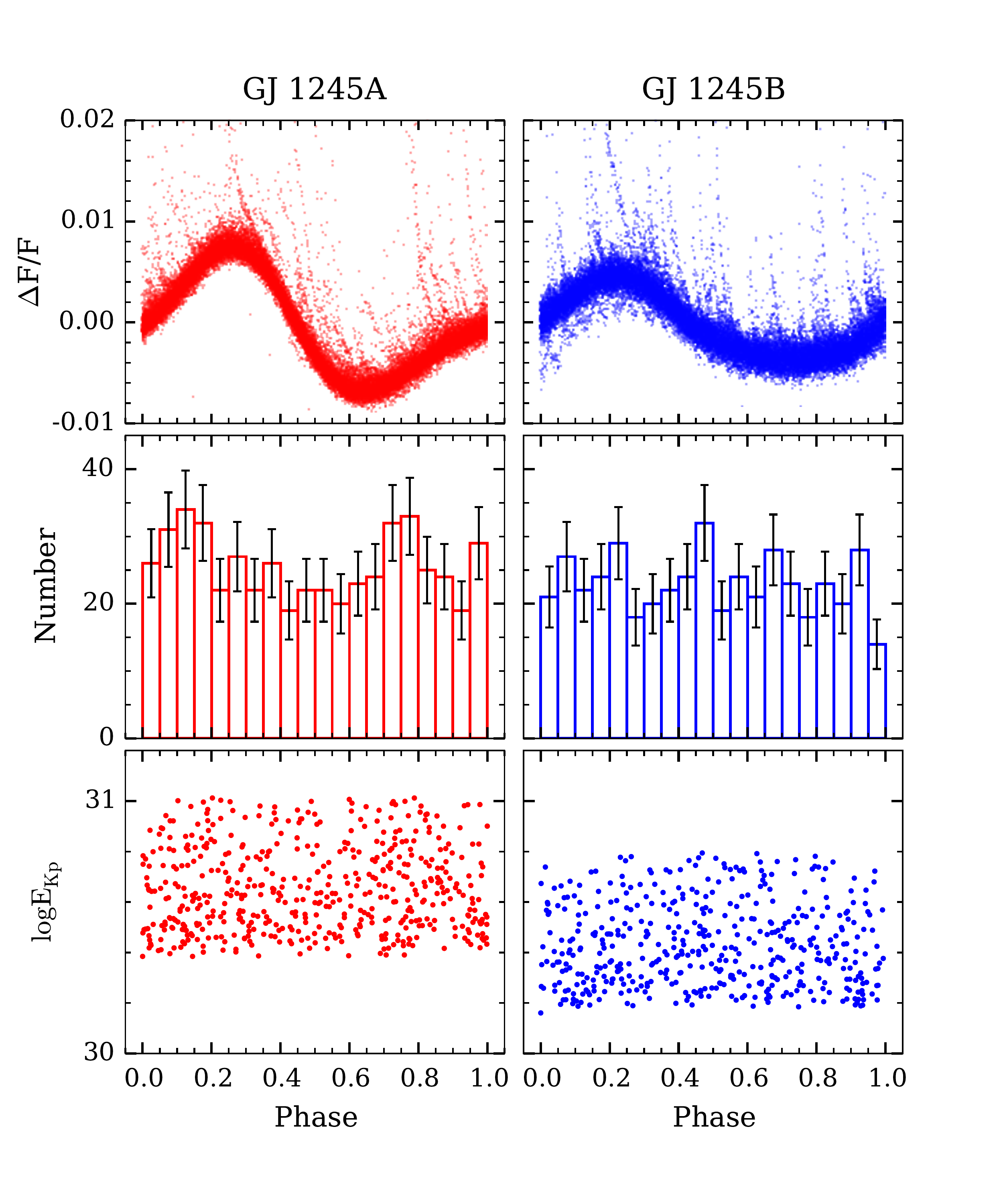}
\caption{In the top panels, one month of short cadence separated light
  curves have been folded at the rotation period of each star. The
  number of flares (middle panels) and flare energy (bottom panels)
  for the 9 month dataset are plotted versus rotation phase. No
  correlations are seen with rotation phase.}
\label{fig:flarephase}
\end{figure}

In Figure \ref{fig:flarephase}, the number of flares and flare
energies occurring on each star are plotted versus rotation
phase. Only flares with equivalent durations less than 10 seconds are
plotted, for which the energies are well constrained. For reference,
one month of the nine month long separated short cadence light curves
have been folded at the rotation period of the respective star. To the
eye, the number distribution on the A component is suggestive of a
correlation with rotation phase. However, any potential phase
dependence is not statistically significant, given the assumed Poisson
errors. The histogram is consistent with a constant distribution at
the median value of the histogram with a reduced $\chi^2$ of 0.87. For
the B component histogram, the reduced $\chi^2$ of constant fit is
0.88. Similarly, the flare energies show no correlation with rotation
phase. These results suggest that the flare-producing regions are
uniformly distributed in longitude across the star.

\subsection{A New Metric for Comparing Flare Rates}

From the power law distribution in Equation 3, an analytical relation
can be obtained for the total energy, $E_{tot}$, released from flares
with energies in the range $E_{0}$ to $E_{1}$. 
\begin{equation}
E_{tot} = \frac{\beta}{2-\alpha} \left(E_{1}^{2-\alpha} - E_{0}^{2-\alpha}\right)
\end{equation}
Alternatively, in terms of equivalent duration, $\mathcal{E}$  
\begin{equation}
\mathcal{E}_{tot} = \frac{E_{tot}}{L_{\mathrm{Kp}}} = \frac{1}{L_{\mathrm{Kp}}}\frac{\beta}{2-\alpha} \left(\mathcal{E}_{1}^{2-\alpha} - \mathcal{E}_{0}^{2-\alpha}\right)
\end{equation}

The constants for these relations are determined from the power law
fit to the FFD. The total luminosity
emitted in flares relative to the total luminosity through the Kepler
bandpass is
\begin{equation}
\frac{L_{fl}}{L_{\mathrm{Kp}}} = \frac{E_{tot}/t_{exp}}{L_{\mathrm{Kp}}}
= \frac{L_{\mathrm{Kp}}\,\mathcal{E}_{tot}/t_{exp}}{L_{\mathrm{Kp}}}
= \frac{\mathcal{E}_{tot}}{t_{exp}}
\end{equation}
where $t_{exp}$ is the total exposure time of the observations. Note
that by expressing the ratio $L_{fl}/L_{\mathrm{Kp}}$ in terms of
equivalent duration, the Kepler luminosity cancels out, removing a
source of uncertainty in comparing $L_{fl}/L_{\mathrm{Kp}}$ for
different stars.

The values of $L_{fl}/L_{\mathrm{Kp}}$ for components A and B reported
in Table \ref{tab:compare} represent an integration of the power law
distribution (Eqn. 5) over the equivalent duration range 2.7 -- 10.0
seconds. These agree well with the values found by simply summing the
equivalent durations of the observed flares over the same range. Note
that in Table 2 of Paper 1, $L_{fl}/L_{\mathrm{Kp}}$ is reported as
$f_E$. The values reported here supersede that of Paper 1, which was
based on the unresolved light curve for components A and B, and was
integrated over a wider equivalent duration range. We caution that due
to the nature of power law distributions, $L_{fl}/L_{\mathrm{Kp}}$
depends on the range of equivalent durations (or energies)
considered. Thus the limits of integration should be reported along
with the values for $L_{fl}/L_{\mathrm{Kp}}$, and taken into account
when comparing to other stars.

\section{Discussion}
\label{sec:discuss}

\begin{deluxetable*}{lcccccc}
\tabletypesize{\small}
\centering
\tablewidth{0pt}
\setlength{\tabcolsep}{0.1in}
\tablecaption{Comparison of Active M Dwarfs}

\tablehead{
\colhead{Star}&     
\colhead{$\log{L_{Kp}}$}&     
\colhead{$P_{rot}$}&
\colhead{$\log{L_{fl}/L_{\mathrm{Kp}}}$$^{\star}$}&
\colhead{$\alpha$$^{\dagger}$}&
\colhead{$\Delta\Omega$}&
\colhead{$\log{L_{\mathrm{H\alpha}}/L_{\mathrm{bol}}}$}\\
\colhead{}                 &
\colhead{(erg s$^{-1}$)}                 &
\colhead{(days)}    &
\colhead{}                 &
\colhead{}                 &
\colhead{(rad day$^{-1}$)}                 &
\colhead{}                 }

\startdata
GJ 1243$^{+}$  & 30.67 & 0.5927 & $-3.78\pm0.01$ & $1.92\pm0.01$ & 0.004 & $-$3.56 \\
GJ 1245A & 30.01 & 0.2632 & $-3.93\pm0.02$ & $1.99\pm0.02$ & 0.0008& $-$4.14\\
GJ 1245B & 29.80 &  0.709 & $-4.00\pm0.02$ & $2.03\pm0.02$ & 0.009 & $-$3.97\\
\enddata
\label{tab:compare}
\tablenotetext{$+$}{The values for GJ 1243 are taken from Papers 1 and 2.}
\tablenotetext{$\star$}{These values represent the integration of the energy
  distribution power law over the equivalent duration range 2.7 -- 10
  seconds.}
\tablenotetext{$\dagger$}{The slope of the cumulative FFD is $1 - \alpha$.}
\end{deluxetable*}

We have used the PyKE PRF modeling routines to produce separate light
curves for two active M dwarfs, GJ 1245 A and B, which were previously
unresolved in the Kepler pipeline processing. Comparison of the model
output to well constrained astrophysical parameters of the system
confirms that we have successfully deconvolved the two stars. The
model recovers the starspot modulations and flares on each star with
minimal cross-contamination. The angular separation of the two stars,
as determined by the PRF models, decreased in a manner consistent with
an astrometric perturbation due to the orbit of the unseen C
component. Unfortunately, the 4 years of Kepler observations only
cover $\sim$25\% of the total orbital period, and are plagued by
significant systematics. A more robust astrometric analysis lies
beyond the scope of this paper. We hope that our results may encourage
others to conduct a more in-depth search for astrometric perturbations
in the Kepler dataset.

Because GJ 1245 A and B are coeval and have similar masses, we are
able to take a holistic view of the dependence of flare occurrence and
differential rotation on rotation rate. This is summarized in Table
\ref{tab:compare}, in which we also include results for the active M4
star GJ 1243. Although we do not know the age of GJ 1243 relative to
the GJ 1245 system, it is a useful comparison star that has been
studied in a similar manner using Kepler data.  The rotation period
for GJ 1243 is taken from Paper 1, while the flare sample is taken
from Paper 2. For consistency, the flare energy distribution slope,
$\alpha$, and the value for $L_{fl}/L_{Kp}$ for GJ 1243 were
determined using the same criteria described in $\S$\ref{sec:flares}
for GJ 1245 A and B. The measure of differential rotation,
$\Delta\Omega$, corresponds to the value for the fastest migrating
spot feature on each star. It represents a lower limit on the shear
between the equator and the poles, expressed in radians per
day. Differential rotation on these three stars will be examined fully
in a forthcoming paper (Davenport et al. 2014, in prep.). Equivalent
widths for H$\alpha$ are reported in Paper 1. These were converted to
$L_{\mathrm{H\alpha}}/L_{\mathrm{bol}}$ via multiplication by the
$\chi$ factor \citep{Walkowicz2004}, which is the ratio between the
continuum flux near H$\alpha$ and the bolometric flux.

Ideally, the flare rates among stars should be compared in terms of
the luminosity emitted in flares relative to the bolometric
luminosity, $L_{fl}/L_{bol}$, similar to the use of
$L_{\mathrm{H\alpha}}/L_{\mathrm{bol}}$ to compare the luminosity
emitted in $H\alpha$. Observationally, we have determined the
luminosity in flares relative to the luminosity in the Kepler
bandpass, $L_{fl}/L_{\mathrm{Kp}}$. Conversion of
$L_{fl}/L_{\mathrm{Kp}}$ to $L_{fl}/L_{bol}$ requires knowing the
color-dependent bolometric correction for the Kepler filter, which is
being developed in a future work (Davenport et al. 2015, in prep.). We
caution that the value of $L_{fl}/L_{\mathrm{Kp}}$, and therefore
$L_{fl}/L_{bol}$, depends on the range of equivalent durations over
which the energy distribution is integrated.  These considerations
must be taken into account when comparing different stars, especially
those that differ significantly in spectral type.

Because GJ 1245 A and B are nearly the same color, we can neglect the
bolometric correction, and compare their values of
$L_{fl}/L_{\mathrm{Kp}}$ as representative of $L_{fl}/L_{bol}$. We
find that GJ 1245A emits a slightly higher fraction of energy in
flares, while Paper 1 found that GJ 1245B emits a slightly
higher fraction of energy in H$\alpha$ emission.  Interestingly, the
values of $L_{fl}/L_{\mathrm{Kp}}$ and
$L_{\mathrm{H\alpha}}/L_{\mathrm{bol}}$ are comparable for the range
of flare equivalent durations we have considered. The scatter in
$L_{\mathrm{H\alpha}}/L_{\mathrm{bol}}$ for stars of the same spectral
type \citep{West2011} easily accounts for the difference between the A
and B components. A similar scatter in $L_{fl}/L_{\mathrm{Kp}}$ is
likely also present. We therefore do not find a correlation of
activity parameters with rotation rate in the GJ 1245 AB system. From
the measured rotation periods, and assuming radii of 0.15 $R_{\odot}$,
GJ 1245 A and B have rotational velocities of 11 and 29 km/s,
respectively. This well above the threshold velocity of $\sim$4 km/s,
where \citet{Mohanty2003} found no correlation between rotation rate
and activity for stars of this spectral type.

If we compare GJ 1245 A and B to GJ 1243, we find that GJ 1243 has the
largest value of $L_{fl}/L_{\mathrm{Kp}}$, with a trend of increasing
$L_{fl}/L_{\mathrm{Kp}}$ with $L_{\mathrm{Kp}}$. \citet{Lacy1976}
found the opposite trend in the $U$ and $B$ bands, although the values
of $L_{fl,U}/L_U$ and $L_{fl,B}/L_B$ were 1 -- 3 orders of magnitude
larger than what we measure in the Kepler bandpass. This underlines
the importance of converting measures of $L_{fl}$ taken in different
bandpasses to $L_{fl}/L_{bol}$. Future work (Davenport et al. 2015, in
prep.) will apply the flare light curve template presented in Paper 2
to understand the changes in relative flare luminosities in different
bandpasses.

The flare energy distributions of all three stars have values of
$\alpha \approx 2$. This is relevant to studies of the Sun, as the
heating of the corona could be attributed to flares if $\alpha > 2$
for the solar flare energy distribution at lower energies
\citep{Schrijver2012}. However, Paper 2 found that the FFD of GJ 1243
had a shallower slope (smaller $\alpha$) at lower energies, and that
fewer low energy flares were observed than predicted by a power law
slope with $\alpha \approx 2$. The flare samples for GJ 1245 A and B
are incomplete below our signal to noise threshold. We therefore
cannot say whether the stars flare less frequently at lower energies
than predicted by the power law, or whether those flares occurred and
were simply not detected due to noise.

For GJ 1245 A and B, the number of flares and flare energies show no
correlation with rotation phase. The same was observed for GJ 1243 in
Paper 1. This is consistent with a scenario where many small flaring
regions are distributed uniformly with longitude, while the long lived
spot features originate from large, axisymmetric poloidal magnetic
fields, as seen in spectropolarimetric Doppler imaging of GJ 1245 B
\citep{Morin2010}.

In agreement with previous observations \citep{CollierCameron2007} and
models \citep{Kuker2011}, the amount of differential rotation
increases with decreasing rotation rate.  The fastest rotator, GJ 1245
A, shows the least differential rotation, and likely rotates as a
nearly solid body. The slowest rotator, GJ 1245 B, shows the greatest
differential rotation, while the differential rotation of GJ 1243 is
intermediate between GJ 1245 A and B. This is among the first
observational constraints placed on the effect of rotation rate on
differential rotation for M dwarfs, and in the case of GJ 1245 A and
B, perhaps the first constraints for objects of the same age.

Our starspot and flare results are among the most detailed for an M
dwarf multiple system, and involved an extensive analysis of the
Kepler target pixel data. We emphasize the importance of the pixel
data, both for the wealth of information they contain, and as a
cautionary example of CCD non-linearity and saturation. Other
investigations of impulsive phenomena using Kepler may encounter
similar effects. In the broader context of stellar activity, these
results contribute significantly to the existing dataset for fully
convective M dwarfs. Together with other results from the Kepler
program, they will help constrain the effects of age and rotation rate
on stellar activity.


\acknowledgements 

This work was supported by Kepler Cycle 2 GO grant NNX11AB71G and
Cycle 3 GO grant NNX12AC79G. JCL acknowledges the support of the
Washington Research Foundation and the University of Washington
Provost's Initiative in Data-Intensive Discovery. JRAD acknowledges
support from NASA ADP grant NNX09AC77G. JRAD and SLH acknowledge
support from NSF grant AST13-11678. SLH acknowledges support from NSF
grant AST08-07205.

This paper includes data collected by the Kepler mission. Funding for
the Kepler mission is provided by the NASA Science Mission
directorate. Some of the data presented in this paper were obtained
from the Mikulski Archive for Space Telescopes (MAST). STScI is
operated by the Association of Universities for Research in Astronomy,
Inc., under NASA contract NAS5-26555. Support for MAST for non-HST
data is provided by the NASA Office of Space Science via grant
NNX13AC07G and by other grants and contracts

This paper made use of PyKE \citep{Still2012}, a software package for
the reduction and analysis of Kepler data. This open source software
project is developed and distributed by the NASA Kepler Guest Observer
Office.  We thank Todd Henry and Fritz Benedict for sharing the
orbital parameters for GJ 1245AC prior to publication. We would like
to acknowledge the observing proposal of Steve Howell, under which the
NOAO archived 2011 observation on the WIYN 0.9m was made.

\bibliographystyle{apj} 
\bibliography{GJ1245.arxiv}


\end{document}